\documentclass{aa}  
\usepackage{graphicx}
\usepackage{txfonts}
\newcommand{\AJ}[3]{{#1}, AJ, \vol{{#2}}, {#3}}
\newcommand{\ApJ}[3]{{#1}, ApJ, \vol{{#2}}, {#3}}
\newcommand{\ApJS}[3]{{#1}, ApJS, \vol{{#2}}, {#3}}

\newcommand{\AnnRev}[3]{{#1}, Annual Review of Astron. and Astrophys.,
\vol{{#2}}, {#3}}
\newcommand{\AandA}[3]{{#1}, A\&A, \vol{{#2}}, {#3}}
\newcommand{\AandAsupp}[3]{{#1}, A\&A Supp.\rm, \vol{{#2}}, {#3}}
\newcommand{\AN}[3]{{#1}, Astron. Nachr.\rm, \vol{{#2}}, {#3}}
\newcommand{\MNRAS}[3]{{#1}, MNRAS\rm, \vol{{#2}}, {#3}}

\newcommand{\PASP}[3]{{#1}, PASP\rm, \vol{{#2}}, {#3}}

\newcommand{\vol}[1]{{\mbox{#1}}}

\newcommand{\Mb}{\mbox{$M_{B}\,$}}
\newcommand{\Mv}{\mbox{$M_{V}\,$}}
\newcommand{\Mk}{\mbox{$M_{K}\,$}}
\newcommand{\Mh}{\mbox{$M_{H}\,$}}
\newcommand{\Mj}{\mbox{$M_{J}\,$}}
\newcommand{\Mi}{\mbox{$M_{I}\,$}}

\newcommand{\Wvi}{\mbox{$W_{VI}\,$}}
\newcommand{\Wjk}{\mbox{$W_{JK}\,$}}

\newcommand{\kms}{\mbox{$\mbox{km\,s}^{-1}$}\,}

\newcommand{\magdex}{\mbox{mag~dex$^{-1}$}\,}

\begin{document}
%

   \title{Calibrating the Cepheid Period-Luminosity relation from the
infrared surface brightness technique}
\subtitle{I. The $p$-factor, the Milky Way relations, and a universal
$K$-band relation\thanks{Table 3 is only available in electronic form
at the CDS via anonymous ftp to cdsarc.u-strasbg.fr (130.79.128.5)
or via http://cdsweb.u-strasbg.fr/cgi-bin/qcat?J/A+A/}}

   \author{
          J. Storm\inst{1}
	  \and
	  W. Gieren\inst{2}
	  \and
	  P. Fouqu\'e\inst{3}
	  \and
          T.G. Barnes\inst{4}
	  \and
          G. Pietrzy\'nski\inst{2,5}
	  \and\\
	  N. Nardetto\inst{2,6}
	  \and
          M. Weber\inst{1}
	  \and
          T. Granzer\inst{1}
	  \and
          K. Strassmeier\inst{1}
          }


   \institute{
             Leibniz-Institut f\"ur Astrophysik Potsdam (AIP),
             An der Sternwarte 16,
             D-14482 Potsdam, Germany,
             \email{jstorm@aip.de}
	 \and
	  Universidad de Concepci\'on, Departamento de Astronom\'ia, 
Casilla 160-C, Concepci\'on, Chile
	 \and
	 IRAP, Universit\'e de Toulouse, CNRS, 
	 14 av. E. Belin, F-31400 Toulouse, France
	 \and
        University of Texas at Austin, McDonald Observatory, 
        82 Mt. Locke Rd, McDonald Observatory, TX 79734 USA
	\and
	Warsaw University Observatory, Al. Ujazdowskie 4, 00-478,
Warsaw, Poland
	\and
	Laboratoire Fizeau, UNS/OCA/CNRS UMR6525, Parc Valrose, 06108
Nice Cedex 2, France
             }

   \date{Received 29 April 2011 / Accepted 23 July 2011}

 
\abstract
{} 
{We determine Period-Luminosity relations for Milky Way Cepheids in
the optical and near-IR bands. These relations can be used directly
as reference for extra-galactic distance determination to Cepheid
populations with solar metallicity, and they form the basis for a
direct comparison with relations obtained in exactly the 
same manner for stars in the Magellanic Clouds, presented in an 
accompanying paper. In that paper we show that the metallicity effect is
very small and consistent with a null effect, particularly in the near-IR bands,
and we combine here all 111 Cepheids from the Milky Way, the
LMC and SMC to form a best relation.}
{We employ the near-IR surface brightness (IRSB) method to determine direct
distances to the individual Cepheids after we have recalibrated
the projection factor using the recent parallax measurements to ten
Galactic Cepheids and the constraint that Cepheid distances to the 
LMC should be independent of pulsation period.}
{We confirm our earlier finding that
the projection factor for converting radial velocity to pulsational
velocity depends quite steeply on pulsation period, $p=1.550-0.186\log(P)$ 
in disagrement with
recent theoretical predictions.  We find PL relations based on 70 
Milky Way fundamental mode Cepheids of $\Mk=-3.33(\pm0.09) (\log(P)-1.0) -
5.66(\pm0.03)$, $\Wvi = -3.26(\pm0.11) (\log(P) - 1.0) - 5.96(\pm0.04)$. 
Combining the 70
Cepheids presented here with the results for 41 Magellanic Cloud
Cepheids which are presented in an accompanying paper,
we find $\Mk=-3.30(\pm0.06)
(\log(P)-1.0) - 5.65(\pm0.02)$, $\Wvi = -3.32(\pm0.08) (\log(P) - 1.0) -
5.92(\pm0.03)$.}
{We delineate the Cepheid PL relation using 111 Cepheids with direct
distances from the IRSB analysis. The relations are by construction in
agreement with the recent HST parallax distances to Cepheids
and slopes are in excellent agreement with the slopes of
apparent magnitudes versus period observed in the LMC.}


\maketitle
%

\section{Introduction}

  In this series of papers we calibrate the Cepheid
Period-Luminosity (PL-) relation using the infrared surface brightness
(IRSB) method.
In Paper~II we address the effect of metallicity on both the slope and
the zero point of the relations in both the optical and near-IR bands
finding very small (consistent with zero) effects in the near-IR and small, 
but possibly significant effects in the optical. 

  Gieren et al. (\cite{Gieren05}) made a first determination of the LMC
PL relations based on thirteen stars with IRSB based distances. 
They found that the distances to the individual Cepheids were dependent 
on the pulsation period which of course is unphysical. They found that
the problem could be resolved by changing the adopted projection ($p$-)
factor relation, which converts observed radial velocities into
pulsation velocities that are needed for Baade-Wesselink type analysis.

  In the present paper we use the new and largely expanded data set 
from Paper~II for now 36 LMC Cepheids together with the new direct 
geometric parallax measurements from Benedict et al. (\cite{Benedict07})
to empirically determine the appropriate $p$-factor relation to be used
in the analyses.

  We present new accurate radial velocity data for 14 galactic
Cepheids expanding the sample to a total of 77 Cepheids, 70 of which can
be used to delineate the Milky Way PL relations. 
We have reanalyzed the complete sample using
exactly the same code and calibrations as for the LMC sample and 
adopting exactly the same reddening law to allow a direct comparison.

  Based on the (near-) universality of the PL relations we combine the 
Milky Way, LMC and SMC samples to determine PL relations based on 111
Cepheids which at the same time constrain the slopes very well, and 
which are tied directly to the parallax zero point from Benedict at al.
(\cite{Benedict07}). These relations thus form a very solid basis 
for the Cepheid distance scale.

The paper is structured as follows: In Sec.\ref{sec.data} we present
the data from the literature as well as new radial velocity data
for fourteen Milky Way Cepheids.
In Sec.\ref{sec.analysis} we present the IRSB method and use the Benedict
et al.  (\cite{Benedict02}, \cite{Benedict07}) parallaxes as well as the
results from Paper~II on 36 LMC Cepheids with IRSB
distances to constrain the $p$-factor relation.  
We then use the new $p$-factor relation to determine distances and
luminosities for 77 Cepheids and derive new PL relations
for the 70 fundamental mode pulsators with good data sets. We proceed
to combine the data with the Magellanic Cloud sample to give our best
global PL relations which can be used for distance determination to
other galaxies. In Sec.\ref{sec.discussion} we compare the results with
other recent investigations and in Sec.\ref{sec.conclusion} we summarize
our conclusions.

\section{The Data}
\label{sec.data}

\begin{table*}
\caption{List of data sources}
\label{tab.dataref}
\centering                          
\begin{tabular}{r c c c | r c c c}
\hline\hline
ID & Optical & $K$-band & rad. vel. &
ID & Optical & $K$-band & rad. vel. \\
\hline
$\eta$~Aql & 4,17 & 3,4 & 9,12,29,35 & U~Nor & 7,16,20 & 2 & 8,21 \\
U~Aql & 17 & 3 & 32 & QZ~Nor & 6,47 & 2 & 28,41 \\
FF~Aql & 13,17 & 3 & 53 & TW~Nor & 16,20 & 2,3,58 & 23,41 \\
FM~Aql & 4,17 & 3,4 & 35 & V340~Nor & 42 & 2 & 23 \\
FN~Aql & 17,20 & 3 & 34 & Y~Oph & 7,17,20 & 2 & 27,55 \\
SZ~Aql & 17,20 & 2 & 35,40 & BF~Oph & 10,17 & 2 & 11,14,15 \\
TT~Aql & 4,7,13,17,20 & 3,4 & 1 & X~Pup & 17,20 & 2 & 1 \\
V496~Aql & 10,17,47 & 3 & 1,48 & AQ~Pup & 7,17 & 2,22,24 & 1 \\
RT~Aur & 13,17 & 4 & 12,34 & BN~Pup & 7,16,20 & 2,22 & 1 \\
$\ell$~Car & 20,38 & 2 & 40 & LS~Pup & 7 & 2,22 & 1 \\
U~Car & 7,20 & 2,3 & 8,40 & RS~Pup & 17,20,47 & 2 & 9 \\
V~Car & 20 & 2 & 15 & VZ~Pup & 7,20 & 2,22 & 1 \\
CT~Car & 51 & 2,22 & 21,41,47 & KQ~Sco & 7 & 2,3 & 8,54 \\
VY~Car & 7 & 2 & 8,15 & RY~Sco & 7,16,17,20 & 2 & 8,14,15 \\
WZ~Car & 7,20 & 2,22 & 8 & EV~Sct & 17,20 & 2 & 9,19,23 \\
SU~Cas & 4,17 & 4 & 9,23 & RU~Sct & 17,20 & 2,3 & 19,41 \\
KN~Cen & 7 & 2 & 8,21 & SS~Sct & 17,20 & 3 & 54 \\
V~Cen & 10,20 & 2 & 11,15 & S~Sge & 4,13,17 & 3,4 & 34 \\
VW~Cen & 7,20 & 2 & 8 & GY~Sge & 43 & 2 & 19,34 \\
XX~Cen & 39 & 2 & 8 & U~Sgr & 10,17 & 2 & 9,23 \\
$\delta$~Cep & 4,17 & 4 & 9,12,23,29,35 & W~Sgr & 17,20 & 3,59,60 & 23 \\
SU~Cru & 20 & 2 & 8,21 & X~Sgr & 17,20 & 3,57 & 1 \\
X~Cyg & 13,17 & 3,4 & 9,29 & Y~Sgr & 17,20 & 3 & 1,56 \\
CD~Cyg & 17 & 3 & 26 & BB~Sgr & 10,17 & 2 & 1 \\
DT~Cyg & 17 & 3 & 23 & WZ~Sgr & 17 & 2,3 & 34 \\
SU~Cyg & 13,17 & 3 & 31 & XX~Sgr & 17 & 3 & 1 \\
VZ~Cyg & 4,17 & 3,4 & 1 & YZ~Sgr & 17,47 & 3 & 1,48 \\
$\beta$~Dor & 20,33 & 2 & 37 & V350~Sgr & 10,17 & 3 & 11,27,34 \\
$\zeta$~Gem & 13 & 57 & 1 & SZ~Tau & 4,17 & 2,4 & 23 \\
X~Lac & 4,17 & 4 & 23 & T~Vel & 47,20 & 2 & 11,15 \\
Y~Lac & 4,17 & 4 & 26 & CS~Vel & 44,45 & 2,22 & 23,41 \\
Z~Lac & 4,17 & 4 & 36 & RY~Vel & 7,16 & 2,3 & 8,21 \\
BG~Lac & 4,17 & 4 & 26 & RZ~Vel & 7,16,20 & 2 & 8,56 \\
GH~Lup & 7,20 & 2 & 8 & SW~Vel & 7,16,20 & 2 & 21,27 \\
T~Mon & 47,7 & 2 & 9,23 & S~Vul & 43 & 2,3 & 34 \\
CV~Mon & 17,20 & 2 & 9,41 & T~Vul & 13,17 & 3,4 & 23 \\
S~Mus & 46 & 2,3 & 30 & U~Vul & 4,13,17 & 4 & 23,25 \\
UU~Mus & 16,39 & 2 & 8 & SV~Vul & 13,17 & 2,3 & 23,26 \\
S~Nor & 44,46 & 2 & 23 &  &  &  &  \\
\hline
\end{tabular}

{\scriptsize
(1) This paper;
(2) Laney \& Stobie, \cite{LS92};
(3) Welch, et al. \cite{Welch84};
(4) Barnes et al. \cite{Barnes97};
(6) Berdnikov \& Turner \cite{BerdnikovTurner95};
(7) Coulson, \& Caldwell \cite{Coulson85a};
(8) Coulson, Caldwell, \& Gieren \cite{Coulson85b};
(9) Storm et al. \cite{Storm04};
(10) Gieren \cite{Gieren81b};
(11) Gieren \cite{Gieren81a};
(12) Kiss \cite{Kiss98b};
(13) Kiss \cite{Kiss98a};
(14) Barnes, Moffett, and Slovak \cite{BMS88};
(15) Lloyd Evans, \cite{Lloyd80};
(16) Madore \cite{Madore75};
(17) Moffett \& Barnes \cite{MB84};
(19) Metzger et al. \cite{Metzger91};
(20) Pel \cite{Pel76};
(21) Pont, Mayor, \& Burki \cite{Pont94};
(22) Schechter et al. \cite{SACK92};
(23) Bersier et al. \cite{Bersier94};
(24) Welch \cite{Welch85};
(25) Imbert \cite{Imbert96};
(26) Imbert \cite{Imbert99};
(27) Petterson et al. \cite{Petterson05};
(28) Kienzle et al., \cite{Kienzle99};
(29) Butler \& Bell \cite{Butler97};
(30) Evans \cite{Evans90a};
(31) Imbert \cite{Imbert84};
(32) Welch et al. \cite{Welch87};
(33) Shobbrook \cite{Shob92};
(34) Gorynya et al. \cite{Gorynya98};
(35) Barnes et al. \cite{Barnes05a};
(36) Sugars \& Evans \cite{SE96};
(37) Wallerstein et al. \cite{Wall92};
(38) Bersier \cite{Bersier02};
(39) Coulson, Caldwell, \& Gieren \cite{Coulson85b};
(40) Bersier \cite{Bersier02};
(41) Metzger et al. \cite{Metzger92};
(42) Bersier et al. \cite{Bers94phot};
(43) Berdnikov \cite{Berd86}, \cite{Berd87}, \cite{Berd92a},
\cite{Berd92b}, \cite{Berd92c}, \cite{Berd92d}, \cite{Berd92e};
(44) Berdnikov \& Turner \cite{BerdnikovTurner98};
(45) Berdnikov  \& Turner \cite{BerdnikovTurner00};
(46) Walraven, Tinbergen, \& Walraven \cite{Walraven64};
(47) Berdnikov \& Caldwell \cite{BerdCald};
(48) Caldwell et al. \cite{Caldwell01};
(51) Pojmanski et al.;
(53) Evans \cite{Evans90b};
(54) Groenewegen \cite{Groenewegen08};
(55) Nardetto et al. \cite{Nardetto06};
(56) Nardetto et al. \cite{Nardetto09};
(57) Feast et al. \cite{Feast08};
(58) McGonegal et al. \cite{McGonegal83};
(59) Kimeswenger et al. \cite{Kimeswenger04};
(60) Wisniewski \& Johnson \cite{Wisniewski68};
}
\end{table*}

We have searched the literature for optical ($V$-band) and near-infrared
($K$-band) light curves as well as radial velocity curves. The starting
point for the search was the catalogues of complete phase coverage
$K$-band light curves for Milky Way Cepheids published by 
Welch et al. (\cite{Welch84}), Laney \& Stobie (\cite{LS92})
and Barnes et al. (\cite{Barnes97}). 
Since the publication of those papers a rich literature of
high quality optical and radial velocity data has materialized and for
the majority of the stars it is now possible to apply the near-infrared surface 
brightness method to determine distances and luminosities. 

In addition to the literature data we have
obtained new, accurate, radial velocity curves for 14 of these
Cepheids to improve the phase coverage and/or data quality for these stars.

We have selected the data sets according to quality and completeness,
but also to ensure, as far as possible, that the data have been obtained
close in time to the near-infrared data to reduce possible errors due to period
variations of the stars. Known double mode pulsators have been
disregarded {\it a priori} as the application of the IRSB method to such stars
could only be attempted if all the data were truly simultaneous.
Thus the list of references reflects this pre-selection and does not
include the data sets that were not used for the analysis.  A number of
first overtone pulsators has also been included, but they are of course not
used for the delineation of the fundamental mode pulsator
Period-Luminosity relations.

In Tab.\ref{tab.dataref} we present the list of stars and the
references to the data sets which we have used in the present analysis. 
A more complete list of of data sets can be found
in Groenewegen (\cite{Groenewegen08}).


The $BVI_c$ photometry reported here is all on the Johnson-Kron-Cousins 
system and the different data sets have been transformed to this system 
when necessary. Similarly all the near-IR data have been transformed 
to the SAAO system as necessary using the transformations from Carter 
(\cite{Carter90}).

The radial velocity data have all been obtained using high
resolution ($R > 20000$) spectrographs. Most of the radial velocities
have been derived using cross-correlation techniques or techniques which
are equivalent. In this way the radial velocities are assumed to be
on a common system and we have not seen indications of significant
systematic differences between datasets for any of our stars for which
we have had data from different techniques. This is an important point
for the application of the IRSB method as the conversion from radial
velocities to pulsational velocities, the so called $p$-factor, depends
to some extent on the procedure which was used for deriving the radial
velocity from the input data.

\subsection{New radial velocity measurements}
\label{sec.STELLArv}

For fourteen of the stars the radial velocity curves were either
missing or they were of limited quality. We have obtained 381 
new radial velocity measurements
for these stars (see Tab.\ref{Tab.STELLAtargets})
using the STELLA Echelle Spectrograph (SES) 
mounted on the fully robotic 1.2m STELLA-I telescope 
(Strassmeier et al. \cite{Strassmeier04}, \cite{Strassmeier10} and 
Weber et al. \cite{Weber08}) at the Iza{\~n}a Observatory on Tenerife, Spain.
SES is a fiber-fed echelle spectrograph with a 2k$\times$2k CCD detector
covering the wavelength range from 388 to 882~nm with small inter-order gaps
starting at 732~nm and increasing towards the red. The resolving power
is $R = 55000$ giving a spectral resolution of 0.12~\AA\, at 650~nm.

\begin{table}
\caption{\label{Tab.STELLAtargets} The Cepheids for which we have
obtained new radial velocity curves. For each star the number of new
data points is listed as well.}
\begin{tabular}{r l | r l}
\hline\hline
Star & $N_{\mbox{\scriptsize obs}}$ & Star & $N_{\mbox{\scriptsize obs}}$ \\
\hline
TT~Aql & 25 & LS~Pup & 26 \\
V496~Aql & 19 & VZ~Pup & 29 \\
$\zeta$~Gem & 65 & X~Sgr & 19 \\
VZ~Cyg & 20 & Y~Sgr & 15 \\
X~Pup & 42 & BB~Sgr & 18 \\
AQ~Pup & 38 & XX~Sgr & 20 \\
BN~Pup & 31 & YZ~Sgr & 14 \\
\hline
\end{tabular}
\end{table}

  The spectra were obtained in fully robotic mode in the period from
March 2007 until July 2010 and reduced using the automatic data reduction 
pipeline (Ritter \& Wassh\"uttl, \cite{RW2004}, Weber et al.
\cite{Weber08}) developed for the instrument. 

\begin{figure*}
\centering
\includegraphics[width=18cm]{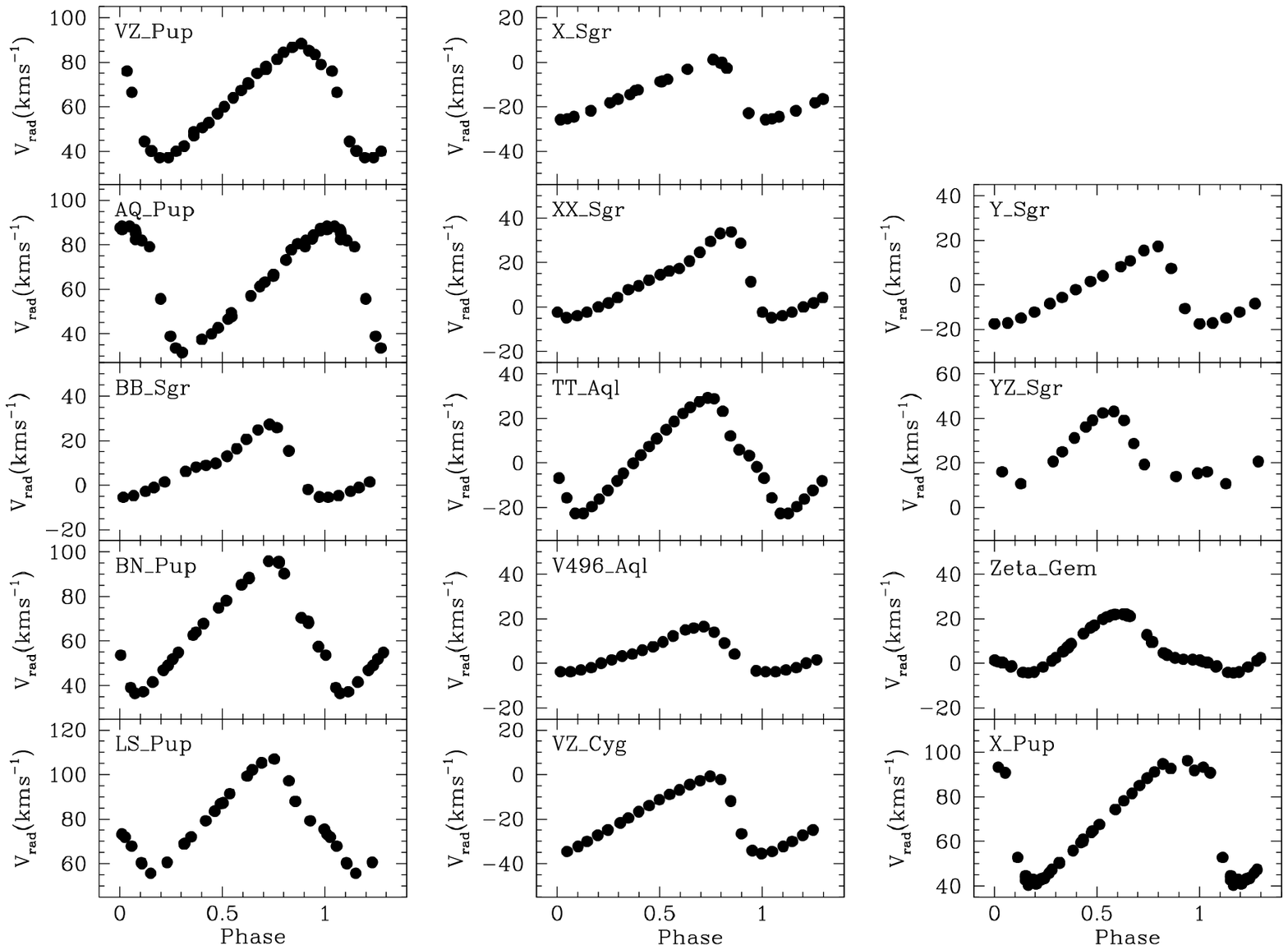}
\caption{\label{Fig.STELLArvall} The new radial velocity curves for fourteen
Milky Way Cepheids from the
STELLA echelle spectrograph as tabulated in Tab.\ref{Tab.STELLArv}.}
\end{figure*}

  The radial velocities returned by the pipeline were corrected for
instrumental velocity offsets and placed on the CORAVEL system by
applying the offset of $+0.503\kms$ determined by 
Strassmeier et al. (in prep). The radial velocities are
tabulated in Tab.\ref{Tab.STELLArv} and plotted in
Fig.\ref{Fig.STELLArvall}.
  
\begin{table}
\caption{\label{Tab.STELLArv}The heliocentric radial velocities (RV) 
measured with the STELLA echelle Spectrograph (SES) as returned by the 
data reduction pipeline and offset to the CORAVEL velocity zero point.
The complete table is available in electronic form from the CDS.}
\begin{tabular}{r r r r r}
\hline\hline
Star & \multicolumn{1}{c}{HJD} & \multicolumn{1}{c}{Phase} &
\multicolumn{1}{c}{RV} & \multicolumn{1}{c}{$\sigma$(RV)} \\
      &    \multicolumn{1}{c}{(Days)}       &      &
\multicolumn{1}{c}{(\kms)} & \multicolumn{1}{c}{(\kms)} \\
\hline
TTAql & 2454175.76609 & 0.53 & 14.79 & 0.04\\
TTAql & 2454213.73669 & 0.29 & -8.15 & 0.06\\
TTAql & 2454218.64326 & 0.65 & 24.95 & 0.07\\
TTAql & 2454222.59739 & 0.94 & 3.25 & 0.02\\
TTAql & 2454223.59473 & 0.01 & -6.92 & 0.02\\
... & ... & ... & ... & ... \\
\hline
\end{tabular}
\end{table}

\subsection{Pulsation velocities}

The $p$-factor (see Sec.\ref{sec.pfactor}), which is used to convert 
the observed radial velocities into pulsational velocities, depends to
some extent on the spectrograph and the procedure used for extracting 
the velocities as different spectral features might carry different
weight in deriving the pulsational velocity.
To investigate this effect for the STELLA velocities we have observed the star
TT~Aql for which an excellent CORAVEL based velocity curve is available
(Imbert \cite{Imbert99}).

\begin{figure}
\centering
\includegraphics[width=9cm]{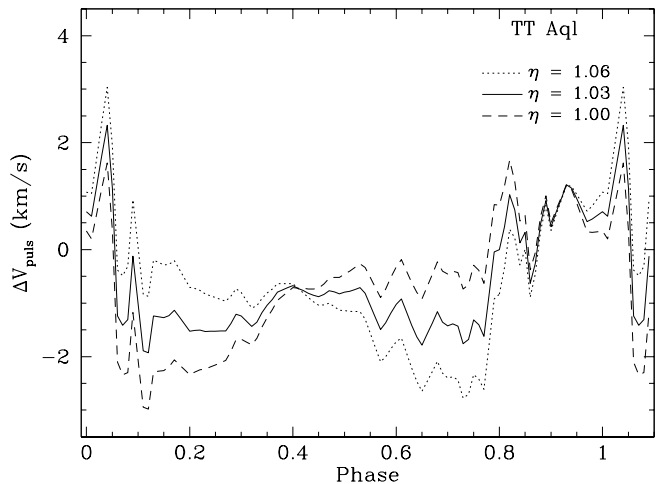}
\caption{\label{Fig.TTAql_drv} The difference in pulsational velocity as a
function of phase between CORAVEL and STELLA measurements for three
different choices of $\eta$ where $\eta=p_{\mbox{\scriptsize
STELLA}}/p_{\mbox{\scriptsize COR}}$.
}
\end{figure}

We follow the procedure described by Storm et al. (\cite{Storm04}) to
determine the ratio between the $p$-factors for STELLA and CORAVEL based
velocities,
$\eta_{\mbox{\scriptsize STELLA}}=p_{\mbox{\scriptsize STELLA}} /
p_{\mbox{\scriptsize COR}}$, for the relevant phase interval, $0 \le
\phi \le 0.8$. In Fig \ref{Fig.TTAql_drv} we plot
the difference in pulsational velocity, 

\begin{equation}
\Delta V_{\mbox{\scriptsize puls}} = 
p_{\mbox{\scriptsize COR}} (\mbox{RV}_{\mbox{\scriptsize COR}} 
- V_\gamma) - 
\eta p_{\mbox{\scriptsize COR}} (\mbox{RV}_{\mbox{\scriptsize STELLA}} 
- V_\gamma)
\end{equation}

\noindent
as a function of phase 
between the linearly interpolated observed radial velocities 
for the two spectrographs for three different values of $\eta$. It
appears that in the phase interval from 0.15 to 0.75 where the velocity
difference curve is smooth, the formally best value is $\eta=1.03$. It
is however also clear that the uncertainty is rather large and given
that the datasets have been obtained at epochs differing by about ten
years, we choose not to apply any additional corrections to the STELLA
velocities but assume that the STELLA and CORAVEL $p$-factors agree to
within 3\%. For the present sample of 14 stars with STELLA velocities
a 3\% effect on the
$p$-factor translates into a 1\% effect on the slopes of the
final PL relations which is much smaller than the statistical errors.

\begin{figure}
\centering
\includegraphics[width=9cm]{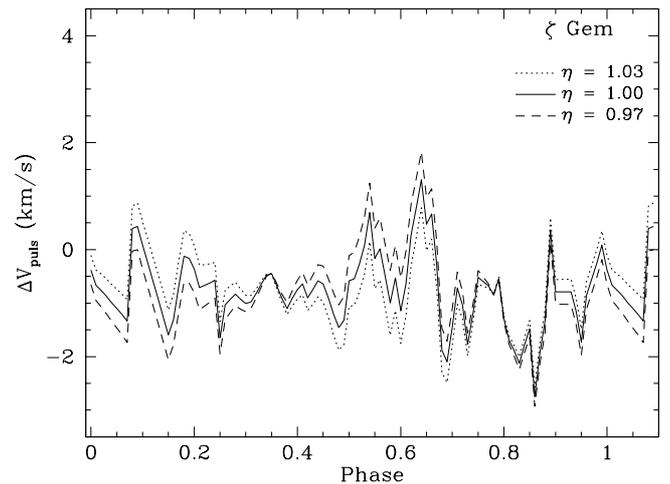}
\caption{\label{Fig.zetaGem_drv} The difference in pulsational velocity as a
function of phase between CORAVEL and HARPS measurements for three
different choices of $\eta$ where $\eta=p_{\mbox{\scriptsize
HARPS}}/p_{\mbox{\scriptsize COR}}$.
}
\end{figure}

  For some stars we have used the radial velocity data from Nardetto et
al. (\cite{Nardetto09}) using the HARPS data pipeline. This data set also
contains data for a star, $\zeta$~Gem, for which a good CORAVEL data
set is available from Bersier (\cite{Bersier94}). Unfortunately the
period of the star is not perfectly constant over time so it has been
necessary to shift the two radial velocity curves with respect to each
other to obtain a good match. Proceeding then as for the STELLA dataset
we find that the two data sets are in good agreement and that
$\eta_{\mbox{\scriptsize HARPS}} = 1.00\pm0.03$ 
as can be seen in Fig.\ref{Fig.zetaGem_drv}. 

  For another star, $\ell$~Car, there is a similar possibility of a
direct comparison between CORAVEL and HARPS data. 
The radial velocity curves from different data sets
exhibit some variations though, and the conclusions are less
straight forward than for $\zeta$~Gem, but they do agree with a value of
$\eta$ which is indistinguishable from unity, which we will adopt in 
the following.

\section{The Analysis}
\label{sec.analysis}

\subsection{The IRSB method}

  The infrared surface-brightness (IRSB) method is a variant of the
Baade-Wesselink method originally developed by Barnes and Evans
(\cite{BarnesEvans76}) in the optical wavelengths. It is based on a functional
relation between a colour index and the surface brightness parameter in
the $V$-band, $F_V$. It was originally calibrated by Welch
(\cite{Welch94}) and a few years later Fouqu\'e and Gieren (\cite{FG97}).
They determined a
very tight linear relationship between the $(V-K)$ colour index
and $F_V$ based on
interferometric angular diameters of giant stars found in the
literature, thus extending the method to the near-infrared. The scatter in
this relation was significantly smaller than was the case for the
optical colour indices used previously. A detailed description of the
implementation of the method which we use in the present paper can be
found in Storm et al. (\cite{Storm04}). 

  Recently direct interferometric angular diameter measurements of
Cepheids have become available (Nordgren et al. \cite{Nordgren02},
Kervella et al. \cite{Kervella04a}, M\'erand et al. \cite{Merand05})
allowing a direct comparison between the surface-brightness relation
for static stars with actual pulsating stars. On this basis Kervella et
al. (\cite{Kervella04b}) find excellent agreement between static and
pulsating stars as well as with the relation determined by Fouqu\'e and Gieren
(\cite{FG97}) for static stars. They find a best fit relation of

\begin{equation}
F_V = -0.1336 (V-K)_0 + 3.9530
\end{equation}

\noindent
with the coefficients determined to better than 2\%.
We adopt their relation for the following analysis.

The surface brightness measure $F_V$ is directly related to the stellar
angular diameter, $\theta$ through the relation

\begin{equation}
\label{eq.Fv}
F_V(\phi) = 4.2207 - 0.1 V_0(\phi) - 0.5 \log \theta(\phi)
\end{equation}
\noindent
where $V_0$ is the de-reddened visual magnitude, and $\phi$ is the phase.

  At the same time geometry gives us the stellar angular diameter
from the stellar radius, $R(\phi)$, through the relation
\begin{equation}
\label{eq.theta}
\theta(\phi) = 2R(\phi) / d  = 2 (R_0 + \Delta R(\phi)) / d
\end{equation}
\noindent
where $\phi$ is the pulsation phase, $d$ is the distance and $R$ is the radius. 

  Integrating the radial velocity curve then gives the radius variation,
$\Delta R(\phi)$ between a reference radius, $R_0$ and the given phase,
$\phi$ as

\begin{equation}
\label{eq.deltaR}
\Delta R(\phi)  = \int -p[V_r(\phi)-V_\gamma]d\phi
\end{equation}
\noindent
where $p$ is the so called projection factor converting radial velocity
into pulsational velocity, $V_r(\phi)$ is the observed radial velocity
and $V_\gamma$ is the systemic velocity.

  We can now solve Eq.\ref{eq.theta} for the two parameters, mean
radius, $R_0$, and distance, $d$ by linear regression to the observed
values of $\theta(\phi)$ from Eq.\ref{eq.Fv} and $\Delta R(\phi)$ from
Eq.\ref{eq.deltaR}. As discussed in Storm et al. (\cite{Storm04}) we use
the OLS bi-sector fit from Isobe et al. (\cite{Isobe90}) for the
regression fit. We fit only the phase interval $\phi \in [0.0,0.8]$ where
the shapes of the two curves usually agrees very well, and we avoid the
phase region $\phi \in [0.8,1.0]$ where the agreement often is quite
poor, most likely due to shocks in the stellar atmosphere.
We also allow for a
small phase shift between the photometric and radial velocity data to
optimize the quality of the fit. The effect on the final PL relation of
these phase shifts is mainly to decrease the scatter in the relation.

\begin{figure}
\centering
\includegraphics[width=9cm]{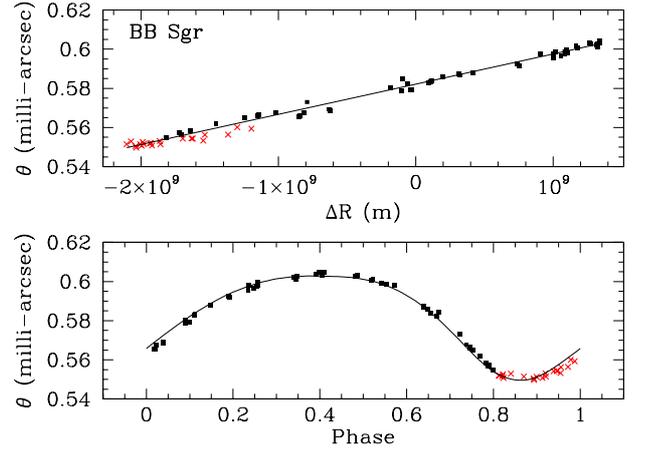}
\caption{\label{Fig.BBSgr} The IRSB fit to the data for the star
BB~Sgr. The deviation between photometric and spectroscopic angular
diameters in the phase interval $\phi \in [0.8,1.0]$ is evident and is
observed for many of the Cepheids in our sample. This
phase interval is therefore disregarded in the fit for all the stars.}
\end{figure}

An example of the fit for the star \object{BB~Sgr} is shown in
Fig.\ref{Fig.BBSgr}. In the upper panel the data used for the actual OLS
bi-sector fit (see Storm et al.  \cite{Storm04} for more details on this)
can be seen and in the lower panel the corresponding photometric angular
diameters have been plotted as filled squares for the points used in
the fit and red crosses for the points in the phase interval $\phi \in
[0.8,1.0]$ which have been disregarded in the fits. The curve in the
lower panel delineates the corresponding spectroscopic angular diameter.

\subsection{Absorption}
\label{sec.absorption}

In order to derive dereddened magnitudes for our Cepheids we use the
colour excess values as given in Fouqu\'e et al. (\cite{Fouque07}). These
values are on the system defined by Laney \& Caldwell (\cite{LC07})
and as discussed by Fouqu\'e et al. (\cite{Fouque07}), these values are
in agreement with the system established by Tammann et al.
(\cite{Tammann03}),
who recalibrated the original measurements compiled by Fernie (\cite{Fernie95}).

For the reddening law we similarly adopt the choice made by Fouqu\'e
et al. (\cite{Fouque07}), namely the law determined by Cardelli et
al. (\cite{Cardelli89}) with a total-to-selective absorption in the $V$
band of $R_V = 3.23$ as determined by Sandage et al. (\cite{Sandage04}).
For the other bands we use $A_I=0.608 \times A_V$, $A_K=0.119 \times A_V$,
$A_J=0.292 \times A_V$, and $A_H=0.181 \times A_V$.

\subsection{Fourier coefficients and identification of fundamental mode pulsators}
\label{sec.fourier}


\begin{figure}
\centering
\includegraphics[width=9cm]{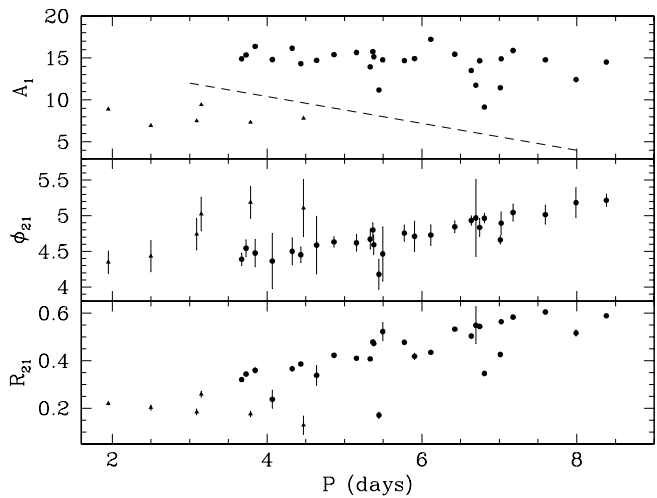}
\caption{\label{Fig.rvfourier} The Fourier parameters based on the
    radial velocity data for the Cepheids with periods less than 9 days.
The dashed line indicates the adopted division between normal Cepheids and
$s$-Cepheids based on Fig.2 in Kienzle et al. (\cite{Kienzle99}). 
Filled circles indicate fundamental pulsators and the triangles, first overtone
pulsators.}
\end{figure}

  Before we attempt to establish the PL relations we have to identify
the fundamental mode pulsators in the sample.  This is difficult
without referring to the period-luminosity diagram.
We use the Fourier parameters for the radial velocity data and the diagrams 
from Kienzle et al. (\cite{Kienzle99}) to reject overtone pulsators. 
To be conservative 
we use the $A_1$ parameter (the radial velocity amplitude) 
to reject $s$-type Cepheids from our final sample as well. In this way we might
remove some bona fide fundamental pulsators as well, but we ensure that
we have a uniform sample. The Fourier parameters for all the Cepheids 
based on the data sets given in Tab.\ref{tab.dataref}
are tabulated in Tab.\ref{tab.fourier}. The error estimates on the
parameters have been calculated using the approximative formula given by
Petersen (\cite{Petersen86}). 

 In Fig.\ref{Fig.rvfourier} the Fourier
parameters for the short period Cepheids have been plotted and the
$s$-Cepheids identified as the stars with $A_1$ below the dashed
line.  These stars are DT~Cyg, EV~Sct, FF~Aql, SU~Cas, SZ~Tau, 
and QZ~Nor. SU~Cas is the shortest-period 
Cepheid in our sample and was probably the first Galactic Cepheid for which
pulsation in the first overtone mode was firmly established by Gieren
(\cite{Gieren76}, \cite{Gieren82}).  Two stars lie below the
sequence of fundamental mode pulsators in the $A_1$ vs. $P$ plot but
above the dashed line and in the other Fourier parameters they appear
unremarkable, so they are not obvious $s$-Cepheids or overtone pulsators.
These stars are X~Lac, and V496~Aql and we will keep them in the sample 
of fundamental mode stars.

\begin{table*}
\caption{\label{tab.fourier}The Fourier parameters for the stars based
on the radial velocity data.}
\scriptsize
\begin{tabular}{r r r r r r r r r r r r}
\hline\hline
ID & Period & Systemic velocity & $A_1$ & $R_{21}$ & $R_{31}$ &
$\phi_{21}$ & $\phi_{31}$ & $\sigma(R_{21})$ & $\sigma(R_{31})$ &
$\sigma(\phi_{21})$ & $\sigma(\phi_{31})$ \\
  & (day) & (\kms) & (\kms) & & & & & & & & \\
\hline
     SU~Cas &   1.94932 & $  -7.36$ &   8.90 &  0.220 &  0.053 &  4.348 &  2.576 &  0.006 &  0.006 &  0.157 &  0.235 \\
     DT~Cyg &   2.49921 & $  -1.70$ &   6.93 &  0.203 &  0.045 &  4.434 &  2.067 &  0.012 &  0.012 &  0.218 &  0.326 \\
     EV~Sct &   3.09099 & $  17.26$ &   7.50 &  0.185 &  0.046 &  4.741 &  2.748 &  0.012 &  0.012 &  0.221 &  0.330 \\
     SZ~Tau &   3.14895 & $  -0.65$ &   9.42 &  0.259 &  0.075 &  5.024 &  3.784 &  0.013 &  0.013 &  0.233 &  0.346 \\
     SS~Sct &   3.67125 & $  -7.24$ &  14.90 &  0.298 &  0.097 &  4.449 &  2.616 &  0.002 &  0.002 &  0.096 &  0.143 \\
     RT~Aur &   3.72832 & $  19.57$ &  15.36 &  0.319 &  0.145 &  4.618 &  3.057 &  0.004 &  0.004 &  0.131 &  0.194 \\
     QZ~Nor &   3.78655 & $ -39.50$ &   7.33 &  0.176 &  0.043 &  5.186 &  4.732 &  0.012 &  0.012 &  0.223 &  1.855 \\
     SU~Cyg &   3.84549 & $   0.08$ &  16.38 &  0.334 &  0.147 &  4.545 &  2.836 &  0.010 &  0.010 &  0.209 &  0.303 \\
     BF~Oph &   4.06751 & $ -28.90$ &  14.81 &  0.222 &  0.137 &  4.420 &  3.110 &  0.035 &  0.035 &  0.423 &  0.580 \\
      Y~Lac &   4.32378 & $ -22.20$ &  16.16 &  0.340 &  0.192 &  4.570 &  2.884 &  0.010 &  0.010 &  0.208 &  0.303 \\
      T~Vul &   4.43541 & $  -2.90$ &  14.32 &  0.358 &  0.162 &  4.517 &  2.829 &  0.004 &  0.004 &  0.123 &  0.185 \\
     FF~Aql &   4.47085 & $  -0.09$ &   7.83 &  0.129 &  0.057 &  5.108 &  3.615 &  0.039 &  0.039 &  0.400 &  0.593 \\
      T~Vel &   4.63982 & $   5.75$ &  14.72 &  0.314 &  0.103 &  4.665 &  3.037 &  0.038 &  0.038 &  0.446 &  0.615 \\
     VZ~Cyg &   4.86445 & $ -18.30$ &  15.39 &  0.392 &  0.155 &  4.716 &  3.302 &  0.002 &  0.002 &  0.082 &  0.123 \\
   V350~Sgr &   5.15424 & $   7.73$ &  15.65 &  0.380 &  0.193 &  4.701 &  3.320 &  0.004 &  0.004 &  0.129 &  0.191 \\
     BG~Lac &   5.33191 & $ -18.10$ &  13.94 &  0.378 &  0.194 &  4.758 &  3.431 &  0.006 &  0.006 &  0.155 &  0.230 \\
$\delta$~Cep &   5.36630 & $ -16.40$ &  15.76 &  0.443 &  0.293 &  4.900 &  3.381 &  0.003 &  0.003 &  0.110 &  0.163 \\
     CV~Mon &   5.37879 & $  19.28$ &  15.15 &  0.437 &  0.212 &  4.672 &  3.219 &  0.005 &  0.005 &  0.152 &  0.225 \\
      X~Lac &   5.44453 & $ -25.20$ &  11.18 &  0.161 &  0.022 &  4.216 &  2.001 &  0.013 &  0.013 &  0.233 &  0.344 \\
      V~Cen &   5.49392 & $ -23.60$ &  14.78 &  0.483 &  0.190 &  4.531 &  2.691 &  0.035 &  0.035 &  0.418 &  0.583 \\
      Y~Sgr &   5.77335 & $  -1.57$ &  14.68 &  0.442 &  0.237 &  4.851 &  3.593 &  0.004 &  0.004 &  0.129 &  0.192 \\
     CS~Vel &   5.90474 & $  26.85$ &  14.93 &  0.388 &  0.171 &  4.802 &  3.522 &  0.013 &  0.013 &  0.237 &  0.346 \\
     FM~Aql &   6.11423 & $  -5.01$ &  17.22 &  0.403 &  0.198 &  4.822 &  3.821 &  0.006 &  0.006 &  0.159 &  0.234 \\
     XX~Sgr &   6.42414 & $  12.65$ &  15.44 &  0.492 &  0.205 &  4.949 &  3.904 &  0.002 &  0.002 &  0.092 &  0.136 \\
     BB~Sgr &   6.63714 & $   8.40$ &  13.50 &  0.466 &  0.152 &  5.047 &  4.346 &  0.001 &  0.001 &  0.076 &  0.113 \\
      V~Car &   6.69668 & $  12.60$ &  11.75 &  0.508 &  0.165 &  5.085 &  4.615 &  0.071 &  0.071 &  0.595 &  0.819 \\
      U~Sgr &   6.74523 & $   2.70$ &  14.67 &  0.503 &  0.190 &  4.939 &  4.005 &  0.005 &  0.005 &  0.139 &  0.206 \\
   V496~Aql &   6.80703 & $   5.10$ &   9.15 &  0.322 &  0.063 &  5.082 &  4.625 &  0.001 &  0.001 &  0.076 &  0.114 \\
      X~Sgr &   7.01283 & $ -12.70$ &  11.45 &  0.395 &  0.148 &  4.748 &  3.459 &  0.001 &  0.001 &  0.065 &  0.098 \\
      U~Aql &   7.02410 & $  -0.11$ &  14.90 &  0.521 &  0.183 &  5.005 &  4.279 &  0.007 &  0.007 &  0.176 &  0.261 \\
$\eta$~Aql &   7.17678 & $ -14.80$ &  15.89 &  0.539 &  0.246 &  5.169 &  4.062 &  0.004 &  0.004 &  0.131 &  0.194 \\
      W~Sgr &   7.59503 & $ -27.70$ &  14.78 &  0.559 &  0.189 &  5.136 &  3.921 &  0.005 &  0.005 &  0.146 &  0.217 \\
      U~Vul &   7.99069 & $   0.24$ &  12.43 &  0.478 &  0.104 &  5.321 &  3.970 &  0.013 &  0.013 &  0.236 &  0.349 \\
      S~Sge &   8.38209 & $ -13.40$ &  14.51 &  0.544 &  0.115 &  5.358 &  4.086 &  0.002 &  0.002 &  0.098 &  0.146 \\
     GH~Lup &   9.27787 & $ -16.10$ &   4.38 &  0.146 &  0.000 &  5.823 &  0.000 &  0.183 &  0.183 &  0.870 & -1.634 \\
     FN~Aql &   9.48151 & $  12.67$ &  14.60 &  0.163 &  0.156 &  0.566 &  5.571 &  0.005 &  0.005 &  0.137 &  0.232 \\
     YZ~Sgr &   9.55369 & $  22.88$ &  13.74 &  0.444 &  0.122 &  0.016 &  4.347 &  0.002 &  0.002 &  0.083 &  0.127 \\
      S~Mus &   9.66007 & $  -0.14$ &  12.85 &  0.479 &  0.137 &  5.820 & -2.260 &  0.011 &  0.011 &  0.209 &  0.338 \\
      S~Nor &   9.75424 & $   5.54$ &  14.31 &  0.332 &  0.161 &  0.255 & -1.444 &  0.007 &  0.007 &  0.168 &  0.373 \\
$\beta$~Dor &   9.84308 & $   8.87$ &  14.01 &  0.313 &  0.192 &  0.337 &  4.914 &  0.017 &  0.017 &  0.263 &  0.626 \\
   $\zeta$~Gem &  10.15073 & $   7.44$ &  11.81 &  0.243 &  0.137 &  0.459 &  5.046 &  0.001 &  0.001 &  0.066 &  0.099 \\
     TW~Nor &  10.78618 & $ -56.60$ &  18.78 &  0.193 &  0.218 &  0.685 &  5.181 &  0.013 &  0.013 &  0.233 &  1.316 \\
      Z~Lac &  10.88564 & $   0.09$ &  18.43 &  0.180 &  0.172 &  0.209 &  5.059 &  0.008 &  0.008 &  0.181 &  0.269 \\
     XX~Cen &  10.95337 & $ -16.20$ &  14.23 &  0.340 &  0.140 &  0.308 &  5.238 &  0.051 &  0.051 &  0.474 &  0.689 \\
   V340~Nor &  11.28700 & $ -40.30$ &   8.58 &  0.265 &  0.053 &  0.206 &  4.441 &  0.016 &  0.016 &  0.413 &  0.391 \\
     UU~Mus &  11.63641 & $ -16.90$ &  20.12 &  0.200 &  0.205 &  0.049 &  5.724 &  0.044 &  0.044 &  0.462 & 69.984 \\
      U~Nor &  12.64371 & $ -22.20$ &  16.88 &  0.135 &  0.087 &  0.206 &  4.921 &  0.044 &  0.044 &  0.448 &  0.646 \\
     SU~Cru &  12.84760 & $ -30.30$ &  11.06 &  0.009 &  0.096 &  3.232 &  5.500 &  0.067 &  0.067 &  1.272 &  0.792 \\
     BN~Pup &  13.67310 & $  67.57$ &  24.86 &  0.206 &  0.021 &  4.495 &  1.650 &  0.001 &  0.001 &  0.069 &  0.103 \\
     TT~Aql &  13.75496 & $   3.83$ &  21.93 &  0.139 &  0.140 &  4.517 &  6.262 &  0.001 &  0.001 &  0.049 &  0.364 \\
     LS~Pup &  14.14640 & $  81.50$ &  22.64 &  0.069 &  0.126 &  6.180 &  5.978 &  0.001 &  0.001 &  0.075 &  0.112 \\
     VW~Cen &  15.03618 & $ -30.60$ &  22.72 &  0.213 &  0.049 &  4.674 &  0.357 &  0.038 &  0.038 &  0.579 &  0.603 \\
      X~Cyg &  16.38633 & $   7.85$ &  22.18 &  0.228 &  0.058 &  4.559 &  0.890 &  0.004 &  0.004 &  0.135 &  0.209 \\
      Y~Oph &  17.12633 & $  -7.12$ &   7.75 &  0.114 &  0.034 &  5.503 &  4.257 &  0.024 &  0.024 &  0.313 &  0.466 \\
     SZ~Aql &  17.14071 & $   6.45$ &  25.38 &  0.263 &  0.039 &  4.563 &  1.735 &  0.003 &  0.003 &  0.112 &  0.167 \\
     CT~Car &  18.05765 & $ 111.10$ &  23.42 &  0.307 &  0.139 &  4.576 &  2.444 &  0.015 &  0.015 &  0.275 &  0.389 \\
     VY~Car &  18.90728 & $   1.86$ &  22.22 &  0.272 &  0.112 &  4.518 &  2.374 &  0.029 &  0.029 &  0.417 &  0.514 \\
     RU~Sct &  19.70062 & $  -5.16$ &  21.14 &  0.244 &  0.014 &  4.753 &  3.027 &  0.009 &  0.009 &  0.236 &  0.330 \\
     RY~Sco &  20.32014 & $ -18.00$ &  15.85 &  0.050 &  0.070 &  4.077 &  4.236 &  0.045 &  0.045 &  0.643 &  0.672 \\
     RZ~Vel &  20.39690 & $  24.81$ &  22.96 &  0.206 &  0.172 &  4.564 &  1.875 &  0.023 &  0.023 &  0.446 &  0.940 \\
     WZ~Sgr &  21.84960 & $ -17.40$ &  23.05 &  0.323 &  0.138 &  4.542 &  2.606 &  0.003 &  0.003 &  0.109 &  0.160 \\
     WZ~Car &  23.01320 & $ -14.40$ &  22.78 &  0.341 &  0.147 &  4.468 &  2.546 &  0.033 &  0.033 &  0.598 &  0.779 \\
     VZ~Pup &  23.17100 & $  63.52$ &  23.00 &  0.303 &  0.121 &  4.446 &  2.450 &  0.001 &  0.001 &  0.068 &  0.088 \\
     SW~Vel &  23.44313 & $  23.75$ &  26.88 &  0.378 &  0.200 &  4.645 &  2.962 &  0.009 &  0.009 &  0.222 &  0.304 \\
      X~Pup &  25.96100 & $  72.02$ &  24.12 &  0.363 &  0.143 &  4.565 &  2.891 &  0.001 &  0.001 &  0.081 &  3.168 \\
      T~Mon &  27.03569 & $  20.67$ &  22.57 &  0.345 &  0.202 &  4.580 &  2.983 &  0.005 &  0.005 &  0.152 &  0.220 \\
     RY~Vel &  28.13117 & $ -10.60$ &  16.18 &  0.267 &  0.127 &  4.333 &  2.464 &  0.039 &  0.039 &  0.600 &  0.602 \\
     KQ~Sco &  28.69580 & $ -30.30$ &  13.84 &  0.940 &  0.611 &  3.943 &  4.083 &  0.038 &  0.038 &  0.628 &  0.616 \\
     AQ~Pup &  30.10400 & $  61.02$ &  23.91 &  0.276 &  0.062 &  4.733 &  3.298 &  0.001 &  0.001 &  0.066 &  0.099 \\
     KN~Cen &  34.02964 & $ -37.20$ &  21.71 &  0.295 &  0.341 &  3.861 &  2.323 &  0.036 &  0.036 &  0.569 &  0.655 \\
$\ell$~Car &  35.54804 & $   2.28$ &  16.77 &  0.296 &  0.150 &  4.639 &  2.949 &  0.005 &  0.005 &  0.153 &  0.225 \\
      U~Car &  38.81233 & $   0.21$ &  21.00 &  0.347 &  0.116 &  4.494 &  2.531 &  0.022 &  0.022 &  0.374 &  0.503 \\
     RS~Pup &  41.47446 & $  25.91$ &  19.81 &  0.444 &  0.192 &  4.615 &  3.103 &  0.008 &  0.008 &  0.185 &  0.267 \\
     SV~Vul &  44.98561 & $  -1.57$ &  19.54 &  0.408 &  0.197 &  4.495 &  3.108 &  0.003 &  0.003 &  0.115 &  0.167 \\
     GY~Sge &  51.56031 & $  15.96$ &  13.15 &  0.301 &  0.104 &  4.766 &  3.118 &  0.014 &  0.014 &  0.250 &  0.362 \\
      S~Vul &  68.77426 & $   1.52$ &  13.26 &  0.335 &  0.208 &  4.742 &  3.323 &  0.008 &  0.008 &  0.181 &  0.268 \\
\hline
\end{tabular}
\end{table*}


\subsection{Constraining the projection factor}
\label{sec.pfactor}

\begin{table}
\caption{\label{tab.HSTresults} The distances for
Cepheids with HST parallax measurements from Benedict et al.
(\cite{Benedict07}). The HST distances are given in column 3, the
IRSB distances using our preferred $p$-factor relation from 
Eq.\ref{eq.pfactor} are given in column 4 and the difference in 
column 5. The uncertainty on the difference is given in the last column.}
\begin{tabular}{r r r r r r}
\hline\hline
(1) & (2) & (3) & (4) & (5) & (6) \\
ID & $\log (P)$ & $d(\mbox{HST})$ & $d(\mbox{IRSB})$ & $\Delta d$ & $\sigma(\Delta d)$ \\
           &           & (pc)     &   (pc)    &  (pc) & (pc) \\
\hline
       FF~Aql &  0.650390 &  355.9 &  369.8 & $  14.0$ & 42.3 \\
       RT~Aur &  0.571489 &  416.7 &  389.0 & $ -27.6$ & 33.0 \\
   $\ell$~Car &  1.550820 &  497.5 &  517.6 & $  20.1$ & 47.9 \\
$\delta$~Cep &  0.729678 &  273.2 &  266.7 & $  -6.5$ & 20.3 \\
  $\beta$~Dor &  0.993131 &  318.5 &  326.6 & $   8.1$ & 22.0 \\
  $\zeta$~Gem &  1.006500 &  359.7 &  385.5 & $  25.8$ & 38.3 \\
W~Sgr$^\dagger$ &  0.880529 &  438.6 &  216.8 & $-221.8$ & 71.1 \\
        X~Sgr &  0.845893 &  333.3 &  322.1 & $ -11.2$ & 28.1 \\
        Y~Sgr &  0.761428 &  469.5 &  436.5 & $ -33.0$ & 75.6 \\
        T~Vul &  0.646934 &  526.3 &  542.0 & $  15.7$ & 59.4 \\
\hline
\end{tabular}

$\dagger$ Not considered in the fits. This star is a known spectroscopic binary.
\end{table}

\begin{figure}
\centering
\includegraphics[width=9cm]{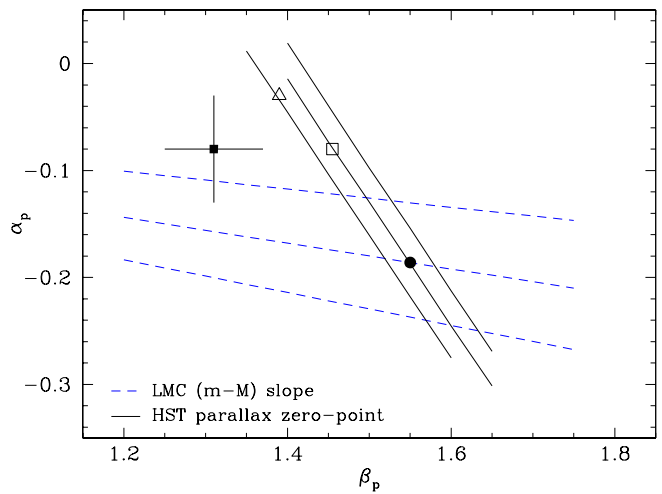}
\caption{\label{Fig.goodpfac} Constraints on the $p$-factor relation
from the HST parallax of Milky Way Cepheids (full line), and from requiring
the LMC distance to be independent of the pulsation period of the stars
(dashed line), see Sec.\ref{sec.pfactor} for details. 
The one sigma contours are also shown with thin lines. The
filled square with error bars shows the theoretical relation from 
Nardetto et al. (\cite{Nardetto09}), the open triangle the Hindsley and Bell
(\cite{HindsleyBell86}) relation and the filled circle the best fit. The
open square shows the theoretical constraint shifted in $\beta$ to
comply with the HST parallax values.}
\end{figure}

  For any Baade-Wesselink type method it is necessary to determine the
pulsational velocity of the surface of the star as this is the velocity
curve which is matched against the angular diameter curve from the
photometry. The same is the case when the angular diameter curve is directly
determined from interferometry. The conversion from the observed radial
velocity to pulsational velocity is commonly parametrized with the projection
factor, $p$.

  The $p$-factor is largely a geometrical correction taking into account
the fact that the radial velocity that we measure is based on light
coming from all points of the hemisphere of the star which is visible to
the observer and not just from the surface element which move along the
line of sight to the star. In fact as discussed by Sabbey et al.
(\cite{Sabbey95}) the $p$-factor depends on the 
temperature of the star, which changes with pulsation period,
as it depends on the limb darkening of the star. 

  The $p$-factor used for converting the observed
radial velocities into pulsational velocities has direct consequences
for the derived distances, as it scales directly with the stellar radius
variation, and is probably the largest source of
systematic error for the method. In the past we (e.g. Gieren et al.
\cite{Gieren93}, Storm et al. \cite{Storm04}, Barnes et al.
\cite{Barnes05b}) have used a relation
with a weak period dependence, $p = 1.39 - 0.03 \log P$, based on
theoretical work by Hindsley and Bell (\cite{HindsleyBell86}).

  Later we (Gieren et al. \cite{Gieren05}) found that the use of
this $p$-factor relation with the IRSB method for LMC Cepheids leads to
an unphysical dependence of the distance modulus on the pulsation
period. We found that a $p$-factor relation of $p=1.58-0.15\log P$
removed this period effect, but the conclusion was not very firm
due to the limited sample of only 13 LMC Cepheids. 

  On the theoretical side, Nardetto et al. (\cite{Nardetto07})
have carefully analyzed individual line profiles from pulsating 
atmosphere models and they also found a relation which was 
steeper than the Hindsley and Bell (\cite{HindsleyBell86}) slope. Nardetto
et al. (\cite{Nardetto09}) generalized this work to match the 
cross-correlation technique used in most observational work on radial
velocities and found a relation 

\begin{equation}
\label{Eq.Nardetto-p}
p = 1.31(\pm0.06) - 0.08(\pm0.05) \log P
\end{equation}

With the sample of 36 LMC Cepheids, presented in Storm et al.
(\cite{Storm11b}, hereinafter Paper~II), covering a wide range
of periods we are in a much better position to constrain the $p$-factor 
relation empirically. Furthermore the recent direct
parallax measurements for ten Milky Way Cepheids by Benedict et
al. (\cite{Benedict02}, \cite{Benedict07}) using the Hubble Space Telescope
Fine Guidance Sensor provide a fundamental set of reference data which
we can use to calibrate the $p$-factor relation.

For three of the ten stars with HST parallaxes (Y~Sgr, X~Sgr, and $\zeta$~Gem) 
we present new radial velocities here, significantly improving the 
available data quality. For one star, W~Sgr, which is a known binary
(Szabados \cite{Szabados03}) the IRSB fit is clearly very poor and we
disregard this star in the further analysis. X~Sgr is known to be
affected by a double shockwave in its atmosphere (Mathias et al.
\cite{Mathias06}) but the IRSB fit looks fine and the agreement with
the HST parallax is excellent so we keep it in the sample.
This leaves us with nine stars in common.
  
As a first step we use the theoretical relation from Nardetto et al.
(\cite{Nardetto09}) in Eq.\ref{Eq.Nardetto-p} and apply the IRSB method to
these nine stars. We find a disappointing difference of $-0.30\pm0.05$~mag
in the distance moduli, the IRSB distances being shorter. Applying the
Nardetto et al. relation to the LMC Cepheids in Paper~II we similarly find
an unlikely result, namely an LMC distance modulus of $18.26\pm0.04$,
much shorter than the canonical value of 18.50. So if we proceed using
first principles, we have a serious conflict with the Benedict et
al. (\cite{Benedict07}) result as well as with most recent works on the
LMC distance that have confined the true distance modulus to a value
between 18.4 and 18.6 (e.g. Pietrzy\'nski et al. \cite{Pietrzynski09};
Szewczyk et al. \cite{Szewczyk08}).

  To reconcile these results we have to conclude either that the
theoretical $p$-factor relation is incorrect due to the lack of some
physics, or that there is a period dependence in the IRSB method which is
not properly accounted for in the current theoretical $p$-factor relation.
We attempt to constrain this effect empirically and parametrize it as
a part of the $p$-factor relation which we retain as having the simple
linear form $p = \alpha_p \times \log (P) + \beta_p$.

  We need to determine two parameters, namely the slope ($\alpha_p$) and
the zero point ($\beta_p$) of the $p$-factor relation. We have two
independent constraints, namely that there should be no systematic dependence
of the LMC Cepheid distances with pulsation period, and we should reproduce, on
average, the Benedict et al. (\cite{Benedict07}) distances.

It turns out that these two constraints are largely orthogonal in the
$(\beta_p,\alpha_p)$ plane as can be seen in Fig.\ref{Fig.goodpfac}.
To determine these parameters we simply apply the IRSB method to each
sample of stars (Milky Way and LMC) for an array of slopes and zero
points for the $p$-factor relation and see where the constraints are
fulfilled. We have varied the zero-point in the range $\beta_p \in
[1.2,1.75]$ in steps of 0.05 and the slope of the relation in the range
$\alpha_p \in [-0.31,0.01]$ in steps of 0.03.

  In Tab.\ref{tab.HSTresults}
we have listed the stars with parallax distances from Benedict
et al. (\cite{Benedict02},\cite{Benedict07}). 
For each pair $(\alpha_p,\beta_p)$
we have determined the IRSB distance, $d(\mbox{IRSB)},$ to these
stars and computed the difference $\Delta(d)=d(\mbox{HST}) -
d(\mbox{IRSB})$. To weight the points independently of distance, we have
normalized the values by dividing by the average distance, $d_{avg} =
(d(\mbox{HST}) + d(\mbox{IRSB}))/2.$, before computing the 
offset $\Delta(d)/d_{avg}$. We then took the mean value of these offsets
and determined the values in the $\beta_p - \alpha_p$ plane where this
mean offset is zero. This is a straight line which is shown in
Fig.\ref{Fig.goodpfac} as a full line with the two thin parallel lines
showing the estimated 1-$\sigma$ interval.

 We then turn to the LMC data set and proceed as for the Milky Way
sample and carry out the IRSB analysis for the same set of
$(\alpha_p,\beta_p)$ values. We then look for the points
where the slope of the LMC distance modulus as a function of $\log (P)$ is
zero. For each individual Cepheid distance we apply the distance modulus 
correction $\Delta(m-M)$ from van der Marel and Cioni (\cite{Marel01})
to correct for the inclination of the LMC disk before determining the
slope. In the $\beta_p,\alpha_p$ plane the resulting constraint is shown as a
dashed line and the two thin dashed lines indicate the estimated 1-$\sigma$
interval. 

The best estimate is:

\begin{equation}
\label{eq.pfactor}
p = 1.550(\pm0.04) - 0.186(\pm0.06) \log P
\end{equation}

\noindent
This is shown as a filled circle in Fig.\ref{Fig.goodpfac}. The $p$-factor
law is even a bit steeper with period than the relation which we found
earlier (Gieren et al. \cite{Gieren05}) and it differs even more from the
recent theoretical relation from Nardetto et al. (\cite{Nardetto09})
which is shown as a filled square in Fig.\ref{Fig.goodpfac}. For
reference the relation from Hindsley and Bell (\cite{HindsleyBell86})
has been plotted as an open triangle in the figure. It agrees within one
$\sigma$ with the HST parallax constraint but not with the constraints
from the period dependence for the LMC sample.  Additionally, the open
square shows a relation where we have adopted the slope $\alpha_p=-0.08$
from Nardetto et al. (\cite{Nardetto09}) but forced the zero point,
$\beta_p$, to give agreement with the HST parallaxes.

\begin{figure}
\centering
\includegraphics[width=9cm]{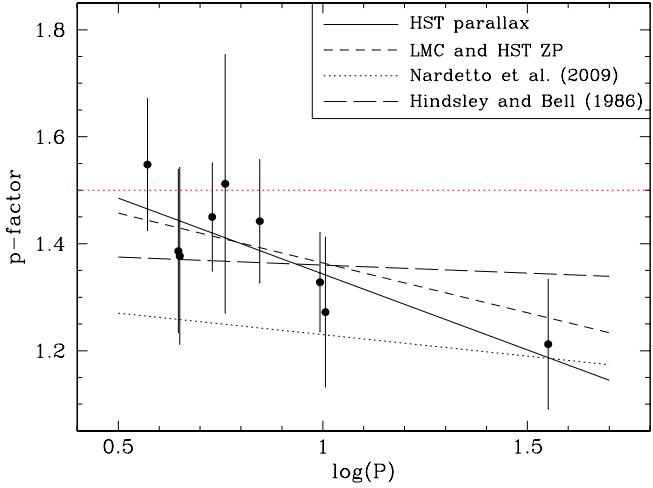}
\caption{\label{fig.pfacHST} The $p$-factor values derived for each of
the HST parallax Cepheids with the only constraint being that the IRSB distance
should agree with the parallax distance. The full line shows the linear
regression fit to the points.  The line labelled "LMC and HST ZP" shows 
our adopted relation based on the constraints from the LMC Cepheids and 
forcing the distance zero point to agree with the HST parallax values. 
The stippled line delineates the theoretical relation from Nardetto et al.
(\cite{Nardetto09}) and the long dashed line the classical Hindsley and
Bell (\cite{HindsleyBell86}) relation. The horizontal line at $p=1.5$
shows the physical limit above which the theoretical $p$-factor would
indicate an unphysical limb-brightening.}
\end{figure}

We can look at the problem in a slightly different way and determine
the $p$-factor for each of the Cepheids with measured HST parallax
distances, and plot them as a function of $\log(P)$ by forcing the IRSB
distance to be equal to the parallax distance.  We have done this and
show the results in Fig.\ref{fig.pfacHST}. A linear fit to these values
gives $p = -0.28(\pm0.08) \log(P) + 1.65(\pm0.07)$. Within the errors
this agrees with our relation (Eq.\ref{eq.pfactor}). We  prefer,
however, to use the relation in Eq.\ref{eq.pfactor}
as it is based on many more stars, especially at pulsation periods
longer than ten days. The linear fit is shown in the figure as well as
our adopted relation (labelled LMC and HST ZP) and the Hindsley and Bell
(\cite{HindsleyBell86}) and Nardetto et al. (\cite{Nardetto09}) relations.  
The horizontal
line at $p=1.5$ indicates the limit above which the $p$-factor would imply
an unphysical limb-brightening instead of the expected limb-darkening. We
note that for the short period stars the $p$-factor is coming close to
this limit.

In Tab.\ref{tab.pfac} we have summarized the resulting values of true 
LMC modulus, the distance offset to the HST parallax data, the slope 
of the LMC Cepheid moduli
as a function of period, for the different assumed $p$-factor relations.
From this Table we can see that the Hindsley and Bell
(\cite{HindsleyBell86}) relation leads to more than 2$\sigma$
deviation for the slope of the LMC distance modulus and thus
seems to be ruled out. The
Nardetto et al. (\cite{Nardetto09}) relation disagrees on both the
constraints, and seems to be ruled out as well. Changing the
value of $\beta_p$ to $\beta_p=1.455$ brings the distance zero point into
agreement with the HST parallax value, but still the distances to the LMC
Cepheids are significantly dependent on the pulsation period. We
thus adopt the fitted relation from Eq.\ref{eq.pfactor} in the
following.

\begin{figure}
\centering
\includegraphics[width=9cm]{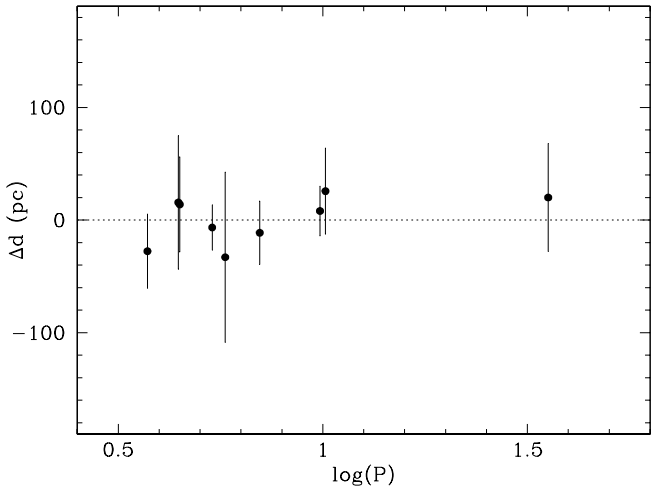}
\caption{\label{fig.HSTdd} The difference between the geometrical
parallax distance measures for nine Cepheids from 
Benedict et al. (\cite{Benedict07}) and the IRSB
distances values, plotted as a function of $\log P$. The revised
$p$-factor relation determined in this paper has been used to calculate
the IRSB distances to the stars.}
\end{figure}

  In Fig.\ref{fig.HSTdd} we have plotted the distance difference between
the HST parallax distances and our IRSB based distance when using the
revised $p$-factor relation. We note that the scatter is very small and
that the data are consistent with no period dependence of the differences.

\begin{table*}
\caption{\label{tab.pfac} The derived quantities for different adopted $p$-factor
relations, $p=\beta_p + \alpha_p \log P$, for the LMC (LMC), and 
Milky Way (MW) samples.  Each column corresponds to a different $p$-factor relation
where the slope and zero point are given in the first two rows.
The PL relations are of the form $M_{m} = a_{m} \times (\log(P) -
1.0) + b_{m}$ where the index $m$ refers to the photometric band. }
\begin{tabular}{r | r r r r }
\hline\hline
$\alpha_{p}$ & $-0.03$ & $-0.08$ & $-0.08$ & $-0.186$ \\
$\beta_{p}$  & 1.39    & 1.31    & 1.455 & 1.550 \\
\hline
Parameter & & & & \\
\hline
$(m-M)_0$(LMC)      & $18.50\pm0.04$ & $18.26\pm0.04$ & $18.50\pm0.04$ & $18.45\pm0.04$\\
 & & & & \\
$(m-M)_0$(LMC) slope & $0.31\pm0.10$ & $0.22\pm0.10$ & $0.24\pm0.10$ & $0.00\pm0.10$\\
$\Delta d$ (pc) & $-5\pm11$ & $-40\pm9$ & $0\pm10$ & $0\pm7$ \\
$\Delta d/d_{avg}$ & $-0.02\pm0.02$ & $-0.11\pm0.02$ & $0.00\pm0.02$ & $0.00\pm0.02$ \\
 & & & & \\
$a_K$  & $-3.58\pm0.09$ & $-3.49\pm0.08$ & $-3.50\pm0.09$ & $-3.33\pm0.09$\\
$b_K$  & $-5.65\pm0.03$ & $-5.45\pm0.03$ & $-5.67\pm0.03$ & $-5.66\pm0.03$\\
 & & & & \\
$a_V$  & $-2.92\pm0.10$ & $-2.83\pm0.10$ & $-2.84\pm0.10$ & $-2.67\pm0.10$\\
$b_V$  & $-3.95\pm0.03$ & $-3.73\pm0.03$ & $-3.97\pm0.03$ & $-3.96\pm0.03$\\
 & & & & \\
$a_{Wvi}$  & $-3.51\pm0.11$ & $-3.42\pm0.12$ & $-3.43\pm0.11$ & $-3.26\pm0.12$\\
$b_{Wvi}$  & $-5.95\pm0.04$ & $-5.74\pm0.04$ & $-5.97\pm0.04$ & $-5.96\pm0.04$\\
 & & & & \\
$a_{Wjk}$  & $-3.69\pm0.10$ & $-3.60\pm0.10$ & $-3.61\pm0.09$ & $-3.44\pm0.09$\\
$b_{Wjk}$  & $-5.95\pm0.03$ & $-5.74\pm0.03$ & $-5.98\pm0.03$ & $-5.96\pm0.03$\\

\hline
\end{tabular}

\end{table*}


%
%
%
%

\subsection{The Period-Luminosity relations}

\begin{table*}
\caption{\label{tab.MWresults}The distances and intensity-averaged 
absolute magnitudes for the complete sample of Milky Way Cepheids based on the IRSB
method as calibrated in this paper. The $\sigma$ values
are the nominal values returned by the bi-sector fitting algorithm. The
Wesenheit indices based on the $(V,I)$ and $(J,K)$ photometry are given
as well as the adopted reddening. The last column gives the adopted phase shift,
$\Delta \phi$, between spectroscopic and photometric data.
}
\scriptsize
\begin{tabular}{r r r r r r r r r r r r r r r r}
\hline\hline
(1) & (2) & (3) & (4) & (5) & (6) & (7) & (8) & (9) & (10) & (11) & (12)
& (13) & (14) & (15) & (16) \\
ID & $\log P$ & $d$ & $\sigma(d)$ & $(m-M)_0$ & $\sigma_{(m-M)}$ & 
$M_B$ & $M_V$ & $M_I$ & $M_J$ & $M_H$ & $M_K$ & $\Wvi$ & $\Wjk$ & $E(B-V)$ & $\Delta \phi$\\
 &  & (pc) & (pc) & (mag) & (mag) & (mag) & (mag) & (mag) & (mag) &
(mag) & (mag) & (mag) & (mag) & (mag) & \\
\hline
\object{     SU Cas} &   0.289884 &    418 &    12 &  8.10 & 0.06 & $-2.53$ & $-2.97$ & $-3.52$ & $-3.82$ & $-4.05$ & $-4.08$ & $-4.37$ & $-4.25$ & 0.259 & $ 0.000$ \\
\object{     DT~Cyg} &   0.397804 &    621 &    30 &  8.97 & 0.11 & $-2.82$ & $-3.32$ & $-3.87$ & $-4.25$ & $-4.51$ & $-4.54$ & $-4.71$ & $-4.75$ & 0.040 & $ 0.030$ \\
\object{     EV Sct} &   0.490098 &   1766 &    82 & 11.24 & 0.10 & $-2.72$ & $-3.21$ & $-3.90$ & $-4.19$ & $-4.45$ & $-4.46$ & $-4.97$ & $-4.64$ & 0.655 & $ 0.045$ \\
\object{     SZ~Tau} &   0.498166 &    558 &     6 &  8.73 & 0.02 & $-2.62$ & $-3.16$ & $-3.80$ & $-4.18$ & $-4.49$ & $-4.53$ & $-4.78$ & $-4.77$ & 0.295 & $-0.005$ \\
\object{     SS~Sct} &   0.564814 &   1000 &    33 & 10.00 & 0.07 & $-2.21$ & $-2.83$ & $-3.55$ & $-3.92$ & $-4.24$ & $-4.27$ & $-4.65$ & $-4.51$ & 0.325 & $-0.035$ \\
\object{     RT~Aur} &   0.571489 &    389 &     4 &  7.95 & 0.02 & $-2.16$ & $-2.69$ & $-3.30$ & $-3.75$ & $-4.01$ & $-4.05$ & $-4.23$ & $-4.26$ & 0.059 & $ 0.005$ \\
\object{     QZ~Nor} &   0.578244 &   1346 &    38 & 10.64 & 0.06 & $-2.00$ & $-2.62$ & $     $ & $-3.76$ & $-4.07$ & $-4.12$ & $     $ & $-4.38$ & 0.263 & $ 0.020$ \\
\object{     SU~Cyg} &   0.584952 &    909 &    21 &  9.79 & 0.05 & $-2.77$ & $-3.24$ & $-3.79$ & $-4.17$ & $-4.45$ & $-4.50$ & $-4.64$ & $-4.73$ & 0.098 & $ 0.000$ \\
\object{     BF Oph} &   0.609329 &    708 &    10 &  9.25 & 0.03 & $-2.04$ & $-2.67$ & $-3.35$ & $-3.77$ & $-4.10$ & $-4.16$ & $-4.39$ & $-4.43$ & 0.235 & $ 0.030$ \\
\object{      Y~Lac} &   0.635863 &   2430 &    35 & 11.93 & 0.03 & $-2.97$ & $-3.48$ & $-4.08$ & $-4.43$ & $-4.74$ & $-4.79$ & $-4.99$ & $-5.04$ & 0.217 & $-0.005$ \\
\object{      T~Vul} &   0.646934 &    543 &     5 &  8.67 & 0.02 & $-2.54$ & $-3.12$ & $-3.74$ & $-4.13$ & $-4.44$ & $-4.50$ & $-4.70$ & $-4.75$ & 0.060 & $ 0.020$ \\
\object{     FF~Aql} &   0.650390 &    369 &    11 &  7.84 & 0.06 & $-2.54$ & $-3.10$ & $-3.72$ & $-4.10$ & $-4.38$ & $-4.42$ & $-4.68$ & $-4.63$ & 0.196 & $-0.030$ \\
\object{      T Vel} &   0.666501 &   1002 &    11 & 10.00 & 0.02 & $-2.28$ & $-2.91$ & $     $ & $-4.05$ & $-4.40$ & $-4.47$ & $     $ & $-4.76$ & 0.289 & $ 0.000$ \\
\object{     VZ~Cyg} &   0.687034 &   1897 &    35 & 11.39 & 0.04 & $-2.68$ & $-3.29$ & $-3.98$ & $-4.36$ & $-4.68$ & $-4.74$ & $-5.04$ & $-5.01$ & 0.266 & $ 0.035$ \\
\object{   V350~Sgr} &   0.712165 &    990 &    23 &  9.98 & 0.05 & $-2.86$ & $-3.47$ & $-4.15$ & $-4.55$ & $-4.89$ & $-4.93$ & $-5.20$ & $-5.19$ & 0.299 & $ 0.000$ \\
\object{     BG~Lac} &   0.726883 &   1776 &    26 & 11.25 & 0.03 & $-2.68$ & $-3.33$ & $-4.05$ & $-4.42$ & $-4.76$ & $-4.82$ & $-5.16$ & $-5.09$ & 0.300 & $ 0.005$ \\
\object{$\delta$ Cep} &   0.729678 &    267 &     5 &  7.13 & 0.04 & $-2.83$ & $-3.41$ & $-4.07$ & $-4.45$ & $-4.78$ & $-4.84$ & $-5.08$ & $-5.11$ & 0.072 & $ 0.000$ \\
\object{     CV Mon} &   0.730685 &   1547 &    24 & 10.95 & 0.03 & $-2.40$ & $-2.97$ & $-3.75$ & $-4.22$ & $-4.58$ & $-4.65$ & $-4.94$ & $-4.94$ & 0.722 & $ 0.020$ \\
\object{      X~Lac} &   0.735997 &   1852 &    38 & 11.34 & 0.04 & $-3.45$ & $-4.02$ & $-4.68$ & $-5.05$ & $-5.38$ & $-5.42$ & $-5.70$ & $-5.68$ & 0.336 & $ 0.040$ \\
\object{      V Cen} &   0.739882 &    643 &    20 &  9.04 & 0.07 & $-2.58$ & $-3.16$ & $-3.82$ & $-4.25$ & $-4.58$ & $-4.65$ & $-4.85$ & $-4.93$ & 0.294 & $ 0.000$ \\
\object{      Y~Sgr} &   0.761428 &    437 &    15 &  8.20 & 0.07 & $-2.40$ & $-3.07$ & $-3.78$ & $-4.25$ & $-4.61$ & $-4.67$ & $-4.87$ & $-4.95$ & 0.188 & $-0.030$ \\
\object{     CS Vel} &   0.771201 &   3184 &   108 & 12.52 & 0.07 & $-2.59$ & $-3.19$ & $     $ & $-4.37$ & $-4.71$ & $-4.78$ & $     $ & $-5.07$ & 0.737 & $ 0.015$ \\
\object{     FM~Aql} &   0.786342 &   1182 &    28 & 10.36 & 0.05 & $-3.31$ & $-4.00$ & $-4.80$ & $-5.12$ & $-5.47$ & $-5.52$ & $-6.05$ & $-5.79$ & 0.589 & $-0.010$ \\
\object{     XX~Sgr} &   0.807815 &   1042 &    42 & 10.09 & 0.09 & $-2.33$ & $-2.92$ & $-3.65$ & $-4.11$ & $-4.44$ & $-4.47$ & $-4.77$ & $-4.73$ & 0.521 & $-0.025$ \\
\object{     BB Sgr} &   0.821971 &    868 &    13 &  9.69 & 0.03 & $-2.96$ & $-3.66$ & $-4.42$ & $-4.86$ & $-5.22$ & $-5.29$ & $-5.60$ & $-5.59$ & 0.281 & $ 0.000$ \\
\object{      V~Car} &   0.825860 &    869 &    16 &  9.69 & 0.04 & $-2.16$ & $-2.86$ & $     $ & $-4.04$ & $-4.40$ & $-4.47$ & $     $ & $-4.77$ & 0.166 & $ 0.000$ \\
\object{      U Sgr} &   0.828997 &    579 &     6 &  8.81 & 0.02 & $-2.73$ & $-3.42$ & $-4.18$ & $-4.61$ & $-4.96$ & $-5.02$ & $-5.34$ & $-5.30$ & 0.402 & $ 0.000$ \\
\object{   V496~Aql} &   0.832958 &    829 &    26 &  9.59 & 0.07 & $-2.37$ & $-3.12$ & $     $ & $-4.32$ & $-4.68$ & $-4.70$ & $     $ & $-4.96$ & 0.397 & $-0.055$ \\
\object{      X~Sgr} &   0.845893 &    322 &     5 &  7.54 & 0.04 & $-3.23$ & $-3.74$ & $-4.35$ & $-4.75$ & $-5.06$ & $-5.11$ & $-5.28$ & $-5.36$ & 0.237 & $ 0.000$ \\
\object{      U~Aql} &   0.846591 &    592 &    19 &  8.86 & 0.07 & $-2.91$ & $-3.58$ & $-4.31$ & $-4.72$ & $-5.06$ & $-5.11$ & $-5.44$ & $-5.38$ & 0.360 & $ 0.015$ \\
\object{    $\eta$ Aql} &   0.855930 &    255 &     4 &  7.03 & 0.09 & $-2.89$ & $-3.55$ & $-4.27$ & $-4.70$ & $-5.04$ & $-5.10$ & $-5.38$ & $-5.38$ & 0.129 & $ 0.000$ \\
\object{      W~Sgr} &   0.880529 &    217 &    16 &  6.68 & 0.16 & $-1.73$ & $-2.37$ & $-3.03$ & $-3.47$ & $-3.80$ & $-3.88$ & $-4.05$ & $-4.17$ & 0.108 & $ 0.040$ \\
\object{      U~Vul} &   0.902584 &    658 &    12 &  9.09 & 0.04 & $-3.22$ & $-3.92$ & $-4.73$ & $-5.03$ & $-5.34$ & $-5.37$ & $-6.00$ & $-5.60$ & 0.603 & $ 0.005$ \\
\object{      S Sge} &   0.923352 &    671 &    16 &  9.13 & 0.05 & $-3.13$ & $-3.84$ & $-4.55$ & $-5.00$ & $-5.34$ & $-5.41$ & $-5.66$ & $-5.70$ & 0.099 & $ 0.000$ \\
\object{     GH~Lup} &   0.967448 &   1307 &    23 & 10.58 & 0.04 & $-3.21$ & $-4.07$ & $-4.90$ & $-5.42$ & $-5.82$ & $-5.90$ & $-6.18$ & $-6.23$ & 0.347 & $-0.040$ \\
\object{     FN~Aql} &   0.976878 &   1175 &    20 & 10.35 & 0.04 & $-2.81$ & $-3.53$ & $-4.36$ & $-4.75$ & $-5.12$ & $-5.17$ & $-5.63$ & $-5.46$ & 0.486 & $-0.030$ \\
\object{     YZ~Sgr} &   0.980171 &   1137 &    13 & 10.28 & 0.03 & $-3.10$ & $-3.84$ & $     $ & $-5.07$ & $-5.44$ & $-5.49$ & $     $ & $-5.77$ & 0.281 & $-0.010$ \\
\object{      S~Mus} &   0.984980 &    858 &    17 &  9.67 & 0.04 & $-3.30$ & $-3.98$ & $     $ & $-5.25$ & $-5.62$ & $-5.71$ & $     $ & $-6.02$ & 0.140 & $ 0.010$ \\
\object{      S Nor} &   0.989194 &    950 &    11 &  9.89 & 0.02 & $-3.27$ & $-4.04$ & $-4.81$ & $-5.33$ & $-5.72$ & $-5.79$ & $-6.01$ & $-6.12$ & 0.179 & $ 0.000$ \\
\object{$\beta$ Dor} &   0.993131 &    327 &     5 &  7.57 & 0.03 & $-3.25$ & $-4.00$ & $-4.73$ & $-5.18$ & $-5.57$ & $-5.63$ & $-5.85$ & $-5.94$ & 0.051 & $ 0.000$ \\
\object{$\zeta$~Gem} &   1.006497 &    386 &     9 &  7.93 & 0.05 & $-3.29$ & $-4.13$ & $     $ & $-5.33$ & $-5.74$ & $-5.81$ & $     $ & $-6.13$ & 0.014 & $ 0.050$ \\
\object{     TW~Nor} &   1.032868 &   2190 &   105 & 11.70 & 0.10 & $-2.96$ & $-3.74$ & $     $ & $-5.29$ & $-5.68$ & $-5.76$ & $     $ & $-6.08$ & 1.157 & $ 0.015$ \\
\object{      Z Lac} &   1.036854 &   1879 &    42 & 11.37 & 0.05 & $-3.42$ & $-4.15$ & $-4.95$ & $-5.37$ & $-5.76$ & $-5.82$ & $-6.18$ & $-6.13$ & 0.370 & $ 0.000$ \\
\object{     XX Cen} &   1.039548 &   1586 &    19 & 11.00 & 0.03 & $-3.35$ & $-4.06$ & $-4.80$ & $-5.26$ & $-5.63$ & $-5.70$ & $-5.94$ & $-6.00$ & 0.271 & $-0.030$ \\
\object{   V340 Nor} &   1.052579 &   1738 &    80 & 11.20 & 0.10 & $     $ & $-3.87$ & $     $ & $-5.23$ & $-5.66$ & $-5.74$ & $     $ & $-6.08$ & 0.322 & $-0.035$ \\
\object{     UU Mus} &   1.065819 &   3146 &   132 & 12.49 & 0.09 & $-3.26$ & $-4.00$ & $-4.79$ & $-5.34$ & $-5.74$ & $-5.82$ & $-6.02$ & $-6.14$ & 0.404 & $ 0.005$ \\
\object{      U Nor} &   1.101875 &   1289 &    39 & 10.55 & 0.07 & $-3.34$ & $-4.08$ & $-4.88$ & $-5.43$ & $-5.82$ & $-5.89$ & $-6.13$ & $-6.21$ & 0.857 & $ 0.000$ \\
\object{     SU~Cru} &   1.108822 &   1274 &    87 & 10.53 & 0.15 & $-3.04$ & $-3.80$ & $     $ & $-5.41$ & $-6.05$ & $-6.15$ & $     $ & $-6.66$ & 0.952 & $ 0.010$ \\
\object{     BN Pup} &   1.135867 &   3817 &    82 & 12.91 & 0.05 & $-3.60$ & $-4.37$ & $-5.17$ & $-5.67$ & $-6.08$ & $-6.15$ & $-6.41$ & $-6.48$ & 0.416 & $ 0.020$ \\
\object{     TT Aql} &   1.138459 &    968 &    15 &  9.93 & 0.03 & $-3.33$ & $-4.20$ & $-5.11$ & $-5.57$ & $-5.98$ & $-6.04$ & $-6.51$ & $-6.36$ & 0.435 & $ 0.005$ \\
\object{     LS Pup} &   1.150646 &   4819 &   113 & 13.41 & 0.05 & $-3.69$ & $-4.46$ & $-5.25$ & $-5.76$ & $-6.17$ & $-6.23$ & $-6.47$ & $-6.56$ & 0.462 & $-0.030$ \\
\object{     VW Cen} &   1.177138 &   3417 &    68 & 12.67 & 0.04 & $-2.94$ & $-3.84$ & $-4.76$ & $-5.43$ & $-5.91$ & $-6.01$ & $-6.19$ & $-6.42$ & 0.439 & $ 0.000$ \\
\object{      X Cyg} &   1.214482 &   1127 &     8 & 10.26 & 0.02 & $-3.69$ & $-4.62$ & $-5.48$ & $-6.00$ & $-6.43$ & $-6.53$ & $-6.82$ & $-6.89$ & 0.228 & $ 0.000$ \\
\object{     CD~Cyg} &   1.232334 &   2427 &    45 & 11.93 & 0.04 & $-3.79$ & $-4.57$ & $-5.45$ & $-5.93$ & $-6.33$ & $-6.40$ & $-6.80$ & $-6.72$ & 0.493 & $ 0.000$ \\
\object{      Y Oph} &   1.233609 &    548 &     6 &  8.70 & 0.03 & $-3.90$ & $-4.62$ & $-5.44$ & $-5.87$ & $-6.21$ & $-6.26$ & $-6.71$ & $-6.53$ & 0.647 & $ 0.020$ \\
\object{     SZ~Aql} &   1.234029 &   2138 &    28 & 11.65 & 0.03 & $-3.88$ & $-4.74$ & $-5.68$ & $-6.20$ & $-6.62$ & $-6.70$ & $-7.13$ & $-7.05$ & 0.534 & $ 0.020$ \\
\object{     CT~Car} &   1.256661 &  10422 &   224 & 15.09 & 0.05 & $-5.32$ & $-4.71$ & $-5.28$ & $-6.16$ & $-6.63$ & $-6.74$ & $-6.17$ & $-7.14$ & 0.570 & $ 0.000$ \\
\object{     VY Car} &   1.276818 &   1728 &    21 & 11.19 & 0.03 & $-3.57$ & $-4.51$ & $-5.38$ & $-5.95$ & $-6.38$ & $-6.47$ & $-6.74$ & $-6.83$ & 0.242 & $ 0.005$ \\
\object{     RU~Sct} &   1.294480 &   1895 &    40 & 11.39 & 0.04 & $-4.12$ & $-4.85$ & $-5.81$ & $-6.47$ & $-6.61$ & $-6.66$ & $-7.29$ & $-6.79$ & 0.911 & $-0.010$ \\
\object{     RY Sco} &   1.307927 &   1128 &    16 & 10.26 & 0.03 & $-3.84$ & $-4.56$ & $-5.42$ & $-5.94$ & $-6.32$ & $-6.39$ & $-6.74$ & $-6.70$ & 0.718 & $ 0.000$ \\
\object{     RZ Vel} &   1.309564 &   1370 &    14 & 10.68 & 0.02 & $-3.74$ & $-4.57$ & $-5.41$ & $-5.99$ & $-6.40$ & $-6.49$ & $-6.71$ & $-6.83$ & 0.300 & $ 0.005$ \\
\object{     WZ Sgr} &   1.339443 &   1774 &    35 & 11.24 & 0.04 & $-3.66$ & $-4.62$ & $-5.60$ & $-6.25$ & $-6.74$ & $-6.84$ & $-7.12$ & $-7.25$ & 0.435 & $-0.015$ \\
\object{     WZ Car} &   1.361977 &   3538 &   108 & 12.74 & 0.07 & $-3.89$ & $-4.68$ & $-5.51$ & $-6.09$ & $-6.50$ & $-6.58$ & $-6.79$ & $-6.92$ & 0.372 & $ 0.000$ \\
\object{     VZ Pup} &   1.364945 &   4591 &    37 & 13.31 & 0.02 & $-4.45$ & $-5.15$ & $-5.91$ & $-6.37$ & $-6.75$ & $-6.80$ & $-7.07$ & $-7.09$ & 0.455 & $-0.075$ \\
\object{     SW Vel} &   1.370016 &   2554 &    34 & 12.04 & 0.03 & $-4.21$ & $-5.02$ & $-5.87$ & $-6.43$ & $-6.85$ & $-6.94$ & $-7.18$ & $-7.29$ & 0.344 & $-0.020$ \\
\object{      X~Pup} &   1.414321 &   2720 &    24 & 12.17 & 0.02 & $-4.26$ & $-5.03$ & $-5.90$ & $-6.39$ & $-6.82$ & $-6.88$ & $-7.24$ & $-7.22$ & 0.421 & $-0.105$ \\
\object{      T Mon} &   1.431915 &   1309 &    19 & 10.59 & 0.03 & $-4.04$ & $-5.04$ & $     $ & $-6.57$ & $-7.04$ & $-7.13$ & $     $ & $-7.52$ & 0.179 & $ 0.000$ \\
\object{     RY Vel} &   1.449158 &   2336 &    36 & 11.84 & 0.03 & $-4.40$ & $-5.23$ & $-6.08$ & $-6.66$ & $-7.04$ & $-7.12$ & $-7.38$ & $-7.44$ & 0.545 & $ 0.000$ \\
\object{     KQ~Sco} &   1.457818 &   2796 &    93 & 12.23 & 0.07 & $-4.16$ & $-5.23$ & $-6.28$ & $-7.03$ & $-7.52$ & $-7.61$ & $-7.91$ & $-8.02$ & 0.869 & $-0.015$ \\
\object{     AQ Pup} &   1.478624 &   3210 &    55 & 12.53 & 0.04 & $-4.65$ & $-5.51$ & $-6.42$ & $-6.92$ & $-7.35$ & $-7.44$ & $-7.82$ & $-7.80$ & 0.518 & $-0.245$ \\
\object{     KN Cen} &   1.531857 &   3586 &    78 & 12.77 & 0.05 & $-4.65$ & $-5.47$ & $-6.33$ & $-7.00$ & $-7.48$ & $-7.58$ & $-7.66$ & $-7.98$ & 0.791 & $ 0.005$ \\
\object{$\ell$~Car} &   1.550816 &    518 &     5 &  8.57 & 0.02 & $-4.18$ & $-5.32$ & $-6.30$ & $-6.94$ & $-7.45$ & $-7.53$ & $-7.83$ & $-7.94$ & 0.146 & $-0.015$ \\
\object{      U~Car} &   1.588970 &   1407 &    17 & 10.74 & 0.03 & $-4.39$ & $-5.30$ & $-6.20$ & $-6.80$ & $-7.23$ & $-7.32$ & $-7.59$ & $-7.68$ & 0.263 & $-0.025$ \\
\object{     RS~Pup} &   1.617420 &   1810 &    30 & 11.29 & 0.04 & $-4.78$ & $-5.76$ & $-6.70$ & $-7.28$ & $-7.73$ & $-7.82$ & $-8.15$ & $-8.19$ & 0.457 & $ 0.035$ \\
\object{     SV~Vul} &   1.652569 &   1895 &    15 & 11.39 & 0.02 & $-4.64$ & $-5.67$ & $-6.63$ & $-7.16$ & $-7.57$ & $-7.62$ & $-8.11$ & $-7.93$ & 0.462 & $ 0.000$ \\
\object{     GY~Sge} &   1.708102 &   2869 &    34 & 12.29 & 0.03 & $-5.39$ & $-6.36$ & $     $ & $-7.80$ & $-8.17$ & $-8.19$ & $     $ & $-8.46$ & 1.310 & $-0.025$ \\
\object{      S~Vul} &   1.837426 &   3762 &    59 & 12.88 & 0.03 & $-5.36$ & $-6.46$ & $     $ & $-8.08$ & $-8.51$ & $-8.57$ & $     $ & $-8.91$ & 0.787 & $ 0.050$ \\
\hline
\end{tabular}
\end{table*}

\begin{figure*}
\centering
\includegraphics[width=18cm]{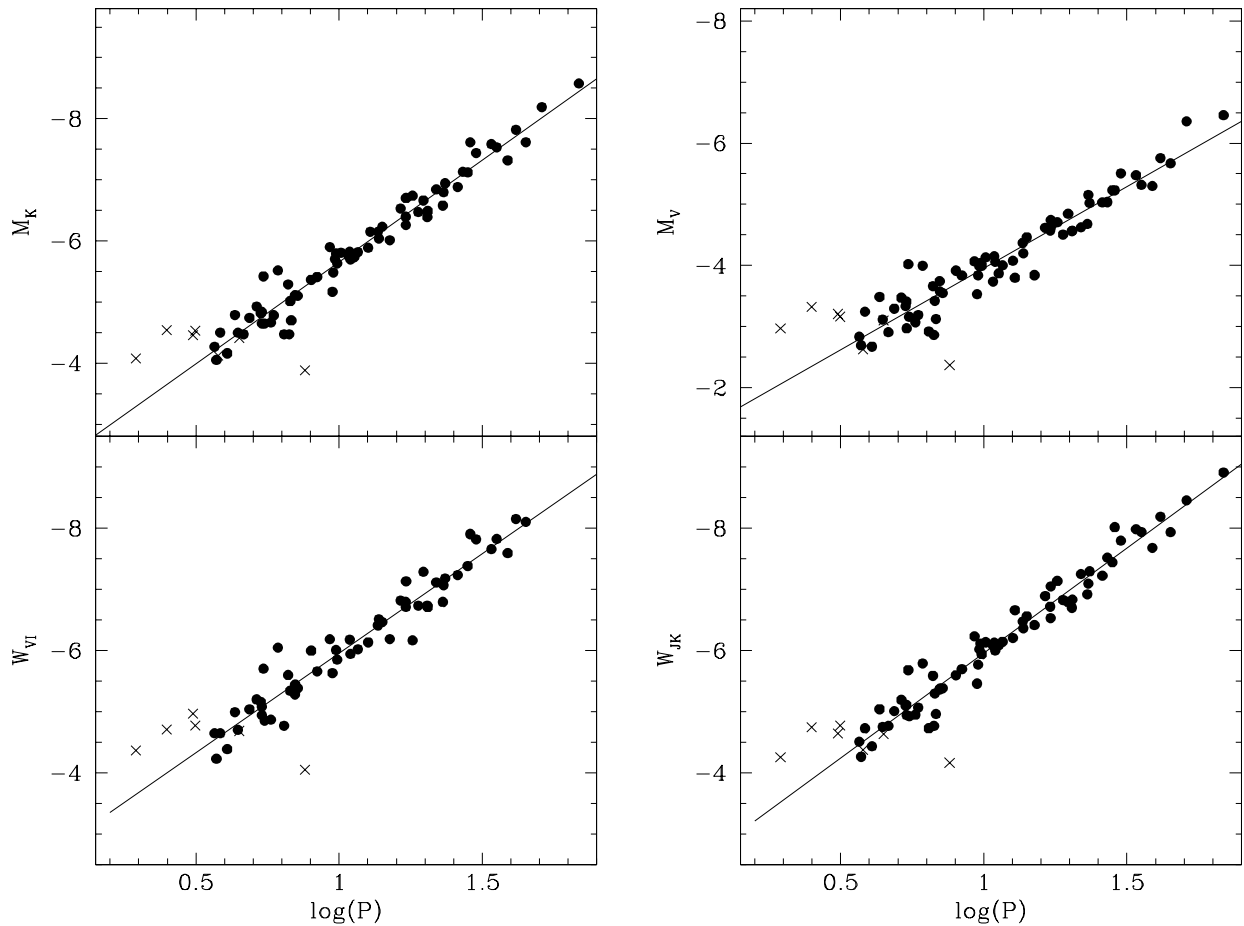}
\caption{\label{Fig.logPMall} The Period-Luminosity relations based on the
IRSB luminosities for our sample of Milky Way Cepheids in the $K$-
and $V$-bands as well as in the Wesenheits indices.
The filled circles represent the fundamental mode Cepheids, and the crosses 
indicate stars which have been disregarded in the linear regression for
reasons mentioned in the text.}
\end{figure*}

\begin{figure}
\centering
\includegraphics[width=9cm]{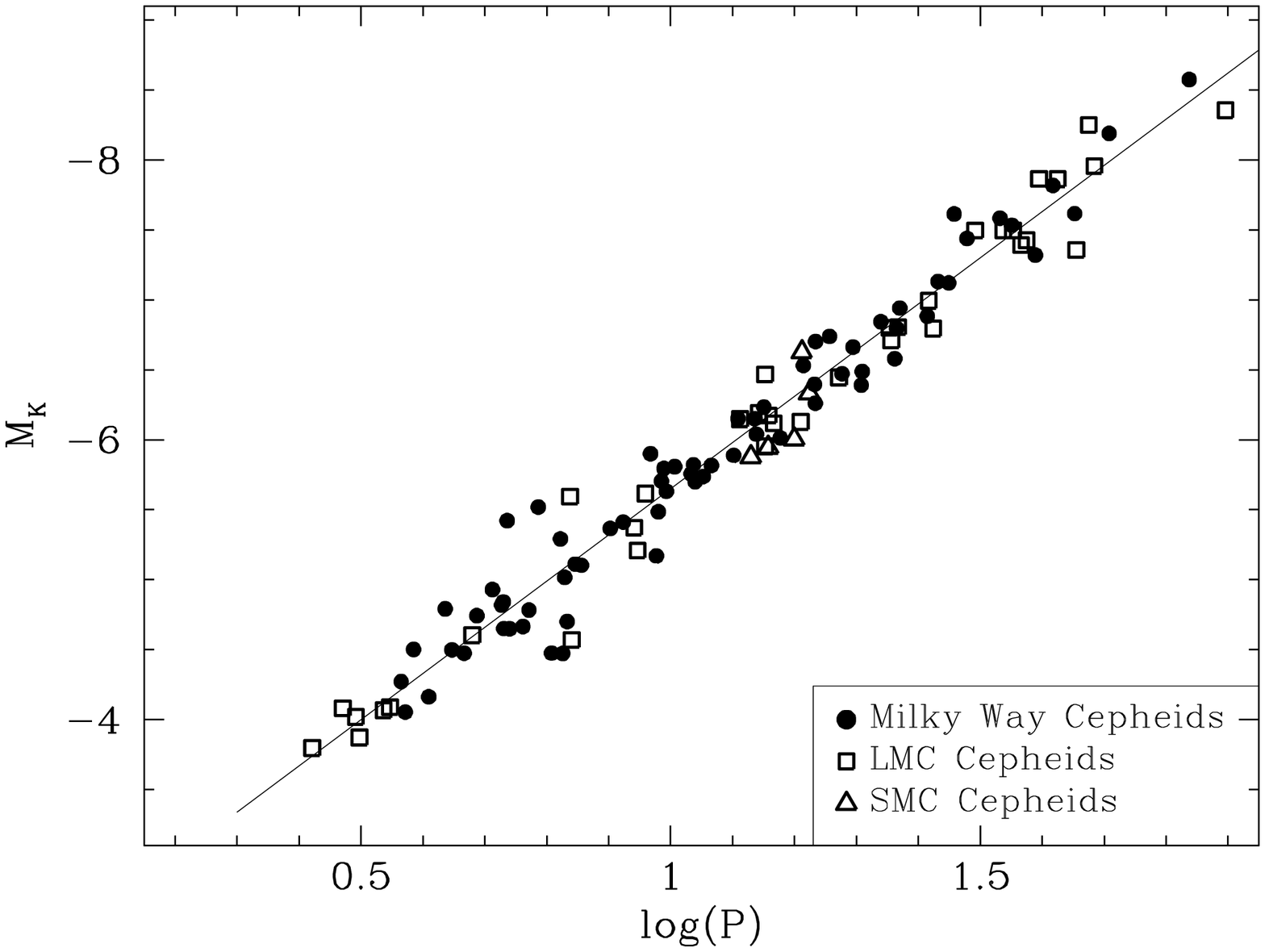}
\caption{\label{Fig.logPMkALL} The Period-Luminosity relation in the
$K$-band for the complete sample of Milky Way, LMC and SMC Cepheids having
IRSB-determined distances in our papers. The overtone pulsators and
stars which have been eliminated for other reasons as discussed earlier
have been eliminated from the plot for clarity.}
\end{figure}

  Using the $p$-factor relation derived in the previous section together with
the reddenings discussed in Sec.\ref{sec.absorption} we obtain the
distances and absolute magnitudes for our Milky Way Cepheids as given in
Tab.\ref{tab.MWresults}. In that Table we present the adopted pulsation
period as well as the distance modulus with the formal uncertainty from
our OLS bi-sector fit (see Storm et al. \cite{Storm04} for details). 
As discussed by Barnes et al. (\cite{Barnes05b})
these uncertainties are underestimated by on average a factor of 3.4 when
compared to the uncertainties returned by the Bayesian fitting technique
employed in that paper.  In the columns 7 to 12
we give the absolute magnitudes in the $B,V,I,J,H,\& K$ bands and in
column 13 and 14 we give the Wesenheit indices (Madore \cite{Madore82})
in the $(V-I)$ band defined as $\Wvi = \Mv - 2.54 (\Mv - \Mi)$ and in the
near-IR $J$ and $K$ band, $\Wjk = \Mk - 0.688 (\Mj-\Mk)$. 
In column 15 we give the adopted reddening
value and in column 16  the adopted phase shift between the
radial velocity data and the photometric data (See Storm et al.
\cite{Storm04} for details).

\begin{table}
\caption{\label{tab.logPM}The Period-Luminosity relations for the
Milky Way Cepheids of the form
$M = a \times (\log(P) - 1.0) + b$. The dispersion around the fit is also tabulated
as well as the formal uncertainties on the coefficients as returned from
the linear regression.}
\begin{tabular}{r c c c}
\hline\hline
Band & $a$ & $b$ & $\sigma$ \\
\hline
\Mb  & $-2.13 \pm 0.13$ & $-3.28 \pm 0.05$ & 0.39 \\ 
\Mv  & $-2.67 \pm 0.10$ & $-3.96 \pm 0.03$ & 0.26 \\ 
\Mi  & $-2.81 \pm 0.10$ & $-4.76 \pm 0.03$ & 0.23 \\ 
\Wvi & $-3.26 \pm 0.11$ & $-5.96 \pm 0.04$ & 0.26 \\ 
\Mj  & $-3.18 \pm 0.09$ & $-5.22 \pm 0.03$ & 0.22 \\ 
\Mh  & $-3.30 \pm 0.08$ & $-5.59 \pm 0.03$ & 0.22 \\ 
\Mk  & $-3.33 \pm 0.09$ & $-5.66 \pm 0.03$ & 0.22 \\ 
\Wjk & $-3.44 \pm 0.09$ & $-5.96 \pm 0.03$ & 0.23 \\ 
\hline
\end{tabular}
\end{table}

Using the data from Tab.\ref{tab.MWresults} for the fundamental mode
pulsators, and after eliminating the binary Cepheid W~Sgr which exhibits a
very poor IRSB fit, we are ready to determine the PL relations in the different bands. In
Tab.\ref{tab.logPM} we list the resulting relations together with
the observed dispersions around the fits. These are our
best estimates of the Milky Way Cepheid PL relations. In
Fig.\ref{Fig.logPMall} we plot the PL relations in the $K$, and $V$ bands
as well as in the Wesenheit indices \Wvi and \Wjk.  We note that the
dispersion around the fits range between 0.22 and 0.39~mag, the $B$-band
relation showing a significantly larger dispersion than the others. For
the other bands the dispersion is only weakly dependent on the
wavelength suggesting that the errors on the absolute magnitudes
are dominated by the distance errors rather than by the intrinsic width
of the PL relation and/or errors in the absorption corrections.
We note that Persson et al. (\cite{Persson04})
found a dispersion of only 0.11~mag for the $K$-band relation in the LMC
whereas in Paper~II we have obtained a value of 0.22~mag, again
suggesting that the dispersion in our current work is dominated by 
distance errors rather than intrinsic luminosity variations between 
Cepheids of similar periods in the sample due to the finite width of 
the instability strip, or errors in the reddenings.

\subsection{The combined sample}

\begin{table}
\caption{\label{tab.PLall}The Period-Luminosity relations based on the 
combined samples of Milky Way, LMC and SMC Cepheids in the form 
$M = a (\log(P) - 1.0) + b$. The metallicity effect, $\gamma$,
on the zero point determined in Paper~II is given in the last column. The
estimated uncertainty on the $\gamma$ value is estimated to be
0.10~\magdex in Paper~II.}
\begin{tabular}{r c c c c}
\hline\hline
\multicolumn{1}{c}{Band}     &  \multicolumn{1}{c}{$a$}   &
\multicolumn{1}{c}{$b$}   &  \multicolumn{1}{c}{Std.dev.} &
\multicolumn{1}{c}{$\gamma$}     \\
     & (\magdex) & (mag) & (mag) & (\magdex) \\
\hline
 \Mv & $-2.73\pm0.07$ & $-3.97\pm0.03$ & 0.26 & $+0.09$\\
 \Mi & $-2.91\pm0.07$ & $-4.75\pm0.02$ & 0.23 & $-0.06$\\
\Wvi & $-3.32\pm0.08$ & $-5.92\pm0.03$ & 0.26 & $-0.23$\\
 \Mj & $-3.19\pm0.06$ & $-5.20\pm0.02$ & 0.22 & $-0.10$\\
 \Mk & $-3.30\pm0.06$ & $-5.65\pm0.02$ & 0.22 & $-0.11$\\
\Wjk & $-3.38\pm0.06$ & $-5.96\pm0.02$ & 0.23 & $-0.10$\\
\hline
\end{tabular}
\end{table}

In Paper~II we show that the PL relations for the Milky Way and
Large Magellanic Cloud samples are identical within the uncertainties,
particularly in the near-IR bands. This means that we can combine the
data from the two papers to derive a PL relation based on a total of
111 Cepheids. In Fig.\ref{Fig.logPMkALL} we have plotted the $K$-band
absolute magnitudes for the Milky Way, LMC and SMC Cepheids from
the two papers together.  The agreement is excellent,
and a linear regression to the combined sample leads to a best
determination for the $K$-band PL relation of:

\begin{equation}
\label{eq.logPMkALL}
\Mk = -3.30(\pm0.06) (\log P - 1.0) - 5.65(\pm0.02)
\end{equation}
\noindent
with a dispersion of 0.22~mag. 
Due to the limited metallicity dependence of this relation found in
Paper~II, this relation can be directly used for distance determination
to galaxies with metallicities between SMC and solar.

We have listed the combined relations in the
other bands, including the Wesenheit indices in Tab.\ref{tab.PLall}, in
all cases without applying any metallicity corrections to the absolute
magnitudes.
The relations have been used in Paper~II to determine the PL relation 
zero-point dependence on metallicity, $\gamma$, and for convenience we have
tabulated those values here as well.

\clearpage

\section{Discussion}
\label{sec.discussion}

\begin{figure}
\centering
\includegraphics[width=9cm]{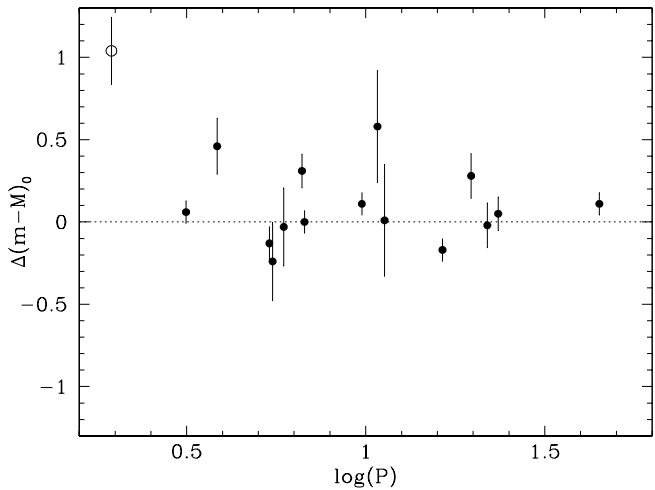}
\caption{\label{fig.TurnerdmM} The difference in derived distance
modulus $\Delta(m-M) = (m-M)_{\mbox{\scriptsize ZAMS}} -
(m-M)_{\mbox{\scriptsize IRSB}}$. The open symbol shows
the star SU~Cas which has been disregarded in the comparison.}
\end{figure}

  The slopes of the PL relations which we find depend directly on the
adopted $p$-factor relation and as we have shown the revised, 
empirically-determined relation is necessary to give pulsation 
period-independent
distances to LMC Cepheids. In Paper~II we further show that the adopted
$p$-factor relation also reproduces the slopes of the apparent magnitude
versus $\log (P)$ relations both in the near-IR and in the optical
bands. The revised $p$-factor relation confirms our earlier findings
(Gieren et al. \cite{Gieren05}) based on a much smaller sample of LMC
Cepheids. Still, it is at odds with the recent careful  theoretical study
by Nardetto et al. (\cite{Nardetto09}). We do not have a ready
explanation for this disagreement but suspect that it is either due to
some missing physics in the theoretical approach or some previously
undetected period dependence in the IRSB method, which we have eliminated
through the parametrization chosen here.
If the problem does not originate from the pulsational velocities, i.e. 
with the $p$-factor, it would have to originate with the 
surface-brightness calibration. The
calibration is based on mostly short period Cepheids with only a single
long period (35~days) Cepheid. However, Kervella et al.
(\cite{Kervella04b}) show a quite convincing comparison of
surface-brightness relations for short and long period stars and the
fact that the relation agrees well with the relation for static stars is
also suggestive that the cause for the steep slope is not buried here. 
The remaining problem seems to be with the short period stars where 
the revised $p$-factor relation leads to values which get close to 1.5, 
suggesting a uniform disk and no limb-darkening. However, the short 
period stars are the ones
which are most numerous in the sample used by Kervella et al.
(\cite{Kervella04b}) and thus the ones which have the best empirical
constraints.  Still we intend to make direct comparisons with
interferometric studies as done by Kervella et al. (\cite{Kervella04c})
for $\ell$~Car in an attempt to understand the reason for the effect which we
see.  For our main purpose of investigating the effect of metallicity on the
Cepheid PL relation by comparing Milky Way and Magellanic Cloud Cepheids 
we are working in a {\em purely differential} way so the actual $p$-factor 
relation cancels out as long as it is the same for both samples of stars, i.e.
metal independent, which from theoretical considerations seems to be a 
good assumption (Nardetto et al. \cite{Nardetto11}).

Groenewegen
(\cite{Groenewegen07}) argues for a constant $p$-factor relation based on a
comparison with a subset of the HST parallax stars but the scatter in
his Fig.2 is very large. With our new high precision radial velocity data for
three of the HST parallax stars, we confirm our $p$-factor
relation (see Fig.\ref{fig.pfacHST}) through
excellent agreement with the full set of Benedict et al.
(\cite{Benedict07}) results on a star by star basis. We also base our
$p$-factor relation on a much larger sample of LMC Cepheids. This sample
has a large
number of stars with pulsation periods significantly longer than 
ten days, thus forming a much firmer basis for constraining the 
$p$-factor relation. In a following paper, 
Groenewegen (\cite{Groenewegen08}) discusses
the use of the Nardetto et al. (\cite{Nardetto07}) relation and finds no
significant difference to a constant value. At the same time he finds
period-luminosity relations for Milky Way Cepheids which are very similar
to the relations presented in the previous section. He finds in the
$K$-band a relation with a slope of $-3.38\pm0.08$, in good agreement with
our value of $-3.33\pm0.09$, similarly he finds a slope in 
the $V$-band of $-2.60\pm0.09$ whereas we find a value of
$-2.67\pm0.10$. It thus seems as if, in spite of the fact that we apply
the same method, there are significant differences in the implementation
of the technique which might affect the results.
  
Recently Molinaro et al. (\cite{Molinaro11}), based on the CORS variant of the
Baade-Wesselink method and using Walraven photometry for 26 galactic
Cepheids and a constant $p$-factor of 1.27, found a PL relation of $\Mv =
-2.78(\pm0.11) \log (P) - 1.42(\pm0.11)$. Again the slope is in good agreement 
with our relation whereas the zero point at a period of ten days differs
by 0.23~mag, our value being fainter.

There are presently very few alternative routes to delineating the 
Milky Way PL relation apart from the Baade-Wesselink type methods. There 
is of course the recent direct parallaxes to ten Cepheids with the HST fine 
guidance sensors by Benedict et al. (\cite{Benedict07}), but this constitutes 
a modest sample of stars for a PL relation. The classical approach is the 
zero-age main sequence (ZAMS) fitting to OB associations and open
clusters containing Cepheids (see e.g. Feast and Walker (\cite{FW87})
and references therein). Turner (\cite{Turner2010}) rederived the
Milky Way PL relation based on the ZAMS fitting to OB associations and
open clusters containing Cepheids finding $\Mv = -2.78(\pm0.12) \log
(P) - 1.29(\pm0.10)$. In Fig.\ref{fig.TurnerdmM} we compare the ZAMS fitting
based distance moduli 
with the IRSB based moduli for the stars in common, and we find very 
good agreement with no significant period dependence. The unweighted 
mean difference is $0.12\pm0.06$~mag so we do find a slight zero point offset. 
We have excluded the star SU~Cas in the comparison as it is an outlier,
even if the IRSB fit appears well defined and does not indicate any
obvious problem with the data.

Benedict et al. (\cite{Benedict07}) found a slope of
$-2.43\pm0.12$ in the $V$ band from the HST parallax measurements, a
value which is only slightly shallower than our value of $-2.67\pm0.10$
and certainly not steeper than our value. 

  We argue that most recent investigations agree to within the errors
with the slope of our relation and they disagree with the earlier
findings of Sandage et al. (\cite{Sandage04}) and Storm et al.
(\cite{Storm04}), that the Milky Way PL relations are significantly steeper
than the LMC relations. In fact, in Paper~II we find that if anything the
optical Milky Way PL relations might be slightly shallower than the 
LMC relations.

  In Paper~II we find that both the slopes and the zero-points of the near-IR 
PL relations are insensitive to metallicity. In addition the $K$-band PL
relation is very insensitive to reddening making this relation
our preferred standard candle. We argue that the best calibration of
 this relation is the combined $K$-band PL relation given in
Tab.\ref{tab.PLall} with the small metallicity effect of 
$\gamma=-0.10\pm0.10$~\magdex also given in that table. We note that for
most extra-galactic Cepheid samples the metallicity is close to the
range from LMC to solar and the failure to correct for the metallicity
effect leads to systematic errors of the order of only 0.02~mag in the
distance modulus. 

  In the optical bands, the LMC and Milky Way slopes are less in
agreement differing by up to 0.2~\magdex as shown in Paper II.
However in the compilation by Bono et al. (\cite{Bono10}) the
slopes in the $V$ and $I$ bands for extra-galactic samples show a large
spread of the order 1~\magdex, much larger than our observed 
difference between the LMC and Milky Way samples. The slope variations 
in that paper do not seem to be strongly correlated with metallicity 
variations so from that point of view our combined SMC, LMC, MW relation 
also provides the better reference relation as it is based on more stars. 

\section{Conclusions and Summary}
\label{sec.conclusion}

  We have obtained new, accurate, radial velocity curves for fourteen Milky Way
Cepheids including three Cepheids with direct parallax measures from
Benedict et al. (\cite{Benedict07}), 
expanding the sample of Milky Way fundamental mode Cepheids to which we
can determine precise IRSB distances to a total of 70 stars.

  We have empirically redetermined the $p$-factor relation, which converts
the observed radial velocities into pulsation velocities needed for the
IRSB method, using two fundamental physical constraints. The first
constraint is that the distance to LMC Cepheids should be independent 
of their pulsation periods, and
the second constraint is that on average we should reproduce the
distances to the Cepheids with parallaxes from Benedict et al.
(\cite{Benedict07}). We find quite a steep relation, $p = 1.550(\pm0.04) -
0.186(\pm0.06) \log (P)$ which is not easily reconciled with recent
theoretical work (e.g Nardetto et al. \cite{Nardetto09}). However, this
revised relation gives rise to PL relations which are in excellent
agreement with other independent determinations both for the Milky Way,
as shown in the present paper, and for the LMC, as shown in Paper~II.

  Using the revised $p$-factor relation we have determined precise PL
relations in the $V,I,J$, \& $K$ bands, as well as the Wesenheit indices
\Wvi, \& \Wjk for these Milky Way stars. These relations can be used for
distance determination to other galaxies with solar abundance.

  In Paper~II we compared these relations to similar relations for a
sample of LMC Cepheids and we found that the effect of
metallicity on the slopes is negligible in the near-IR and small,
possibly consistent with zero, in the
optical bands as well. Including also a sample of SMC Cepheids we find
that the zero points of the PL relations depend on metallicity to a
varying degree, but in most bands the effect is small, of the order
$-0.10\pm0.10$~\magdex, which is consistent with a zero effect. Consequently
we argue that it is warranted to combine the three samples of Cepheids 
giving us a total sample of 111 Cepheids with IRSB distances which can be used
to delineate accurate absolute, universal PL relations.

  Our best standard candle is the $K$-band PL relation as it is not only
insensitive to reddening and shows a low intrinsic dispersion, but it also
exhibits no metallicity dependence on the slope and only a weak
dependence on the zero point consistent with a null effect.
The $K$-band relation based on the full
sample of Cepheids presented here is $\Mk = -3.30(\pm0.06)[\log (P) -
1.0] - 5.65(\pm0.02)$. The combined relation in the optical \Wvi index is $\Wvi
= -3.32(\pm0.08)[\log (P) -1.0] - 5.92(\pm0.03)$.

\begin{acknowledgements}

This research has made use of the SIMBAD database,
operated at CDS, Strasbourg, France.

STELLA is funded by AIP through the State of Brandenburg and the Federal
Ministry for Education and Science in Germany. It is operated jointly by
AIP and the Instituto de Astrofisica de Canarias (IAC) at the Teide
Observatory of the IAC. The authors thank the staff of the IAC for their
great and continuous support to run STELLA.

WG and GP gratefully acknowledge financial support for this work from the
Chilean Center for Astrophysics FONDAP 15010003, and from the BASAL Centro
de Astrofisica y Tecnologias Afines (CATA) PFB-06/2007.

\end{acknowledgements}

%
%

\end{document}